\documentclass[journal]{IEEEtran}
\IEEEoverridecommandlockouts
\usepackage{cite}
\usepackage{amsmath,amssymb,amsfonts}
\usepackage{soul}
\usepackage{algorithm}
\usepackage{algpseudocode}
\usepackage{graphicx}
\usepackage{textcomp}
\usepackage[dvipsnames]{xcolor}
\usepackage{pgfplots,tikz}
\usepackage[font=small]{caption}
\usepackage{subcaption}
\usepackage{comment}
\usepackage{bm}
\usepackage{mathtools}
\def\BibTeX{{\rm B\kern-.05em{\sc i\kern-.025em b}\kern-.08em
    T\kern-.1667em\lower.7ex\hbox{E}\kern-.125emX}}

\newcommand{\calA}{{\cal A}}
\newcommand{\calB}{{\cal B}}
\newcommand{\calC}{{\cal C}}
\newcommand{\calCN}{{\cal CN}}
\newcommand{\calD}{{\cal D}}
\newcommand{\calE}{{\cal E}}
\newcommand{\calK}{{\cal K}}
\newcommand{\calH}{{\cal H}}
\newcommand{\calL}{{\cal L}}
\newcommand{\calM}{{\cal M}}
\newcommand{\calN}{{\cal N}}
\newcommand{\calP}{{\cal P}}
\newcommand{\calS}{{\cal S}}
\newcommand{\calU}{{\cal U}}

\newcommand{\ds}{\displaystyle}

\newcommand{\ba}{\mathbf{a}}
\newcommand{\bb}{\mathbf{b}}
\newcommand{\bd}{\mathbf{d}}
\newcommand{\bg}{\mathbf{g}}
\newcommand{\bh}{\mathbf{h}}
\newcommand{\bp}{\mathbf{p}}

\newcommand{\bs}{\mathbf{s}}
\newcommand{\bw}{\mathbf{w}}
\newcommand{\bx}{\mathbf{x}}
\newcommand{\by}{\mathbf{y}}
\newcommand{\bz}{\mathbf{z}}

\newcommand{\bzero}{\mathbf{0}}

\newcommand{\bA}{\mathbf{A}}

\newcommand{\bC}{\mathbf{C}}
\newcommand{\bD}{\mathbf{D}}

\newcommand{\bF}{\mathbf{F}}
\newcommand{\bG}{\mathbf{G}}
\newcommand{\bH}{\mathbf{H}}
\newcommand{\bI}{\mathbf{I}}
\newcommand{\bP}{\mathbf{P}}
\newcommand{\bQ}{\mathbf{Q}}
\newcommand{\bR}{\mathbf{R}}
\newcommand{\bS}{\mathbf{S}}
\newcommand{\bU}{\mathbf{U}}
\newcommand{\bV}{\mathbf{V}}
\newcommand{\bW}{\mathbf{W}}
\newcommand{\bX}{\mathbf{X}}
\newcommand{\bY}{\mathbf{Y}}
\newcommand{\bZ}{\mathbf{Z}}

\newcommand{\bbE}{\mathbb{E}}

\newcommand{\test}{{\underset{H_0}{\overset{H_1}\gtrless}}}
\newcommand{\norm}[1]{\left\lVert#1\right\rVert}

\usetikzlibrary{spy}
\usetikzlibrary{external}
\pgfplotsset{compat=1.18}

\begin{document}

\title{Scalable Integrated Sensing and Communications for Multi-Target Detection and Tracking in Cell-Free Massive MIMO: A Unified Framework}
\author{
Sergi Liesegang~\IEEEmembership{Member,~IEEE,} Stefano~Buzzi,~\IEEEmembership{Senior Member,~IEEE,} and 
Carmen~D'Andrea,~\IEEEmembership{Member,~IEEE,}
    \thanks{A preliminary and compressed version of this paper has been presented in the conference paper \cite{buzzi2024scalable}.}
    \thanks{The European Union (EU) supported the work of S. Liesegang under the MSCA Postdoctoral Fellowship DIRACFEC (grant agreement No. 101108043). The EU also supported the work of S. Buzzi and C. D'Andrea under the Italian National Recovery and Resilience Plan (NRRP) of NextGenerationEU, partnership on “Telecommunications of the Future” (PE00000001 - program “RESTART”, Structural Project 6GWINET, Cascade Call SPARKS). Views and opinions expressed are those of the authors and do not necessarily reflect those of the EU. The EU cannot be held responsible for them.
    }
    \thanks{
    The authors are with the Department	of Electrical and Information Engineering (DIEI), University of Cassino and Southern Latium (UNICAS), 03043 Cassino, Italy, and the \textit{Consorzio Nazionale Interuniversitario per le Telecomunicazioni} (CNIT), 43124 Parma, Italy. S. Buzzi is also affiliated with the Department of Electronics, Information, and Bioengineering (DEIB), Politecnico di Milano (PoliMI), 20122 Milano, Italy.}
    
}

\maketitle

\begin{abstract}
This paper investigates a cell-free massive MIMO (multiple-input multiple-output) system where distributed access points (APs) perform integrated sensing and communications (ISAC) tasks, enabling simultaneous user communication and target detection/tracking. A unified framework and signal model are developed for the detection of potential targets and tracking of previously detected ones, even in arbitrary positions. Leveraging the Generalized Likelihood Ratio Test technique, novel detection/tracking algorithms are proposed to handle unknown target responses and interference. Scalable AP-user and AP-target association rules are evaluated, explicitly considering multi-zone sensing scenarios. Additionally, a scalable power control mechanism extends fractional power control principles to ISAC, balancing power allocation between communication and sensing tasks. For benchmarking, a non-scalable power control optimization problem is also formulated to maximize the minimum user data rate while ensuring a Quality of Service constraint for sensing, solved via successive convex approximation. Extensive numerical results validate the proposed framework, demonstrating its effectiveness in both communication and sensing, revealing the impact of interference from other targets, and highlighting fundamental trade-offs between sensing and communication performance.
\end{abstract}


\section{Introduction} \label{sec:1}
6G wireless networks will integrate sensing functions into their architecture, enabling integrated sensing and communications (ISAC) \cite{liu2022integrated}. ISAC leverages the dual functionality of wireless signals for both data transmission and environmental awareness, supporting applications like autonomous vehicles, smart cities, and industrial automation. By using the same transceiver hardware and frequency bands for both tasks, ISAC transforms communication infrastructure into a platform for environmental monitoring and object tracking, unlocking new applications and revenue opportunities.

The deployment of ISAC on a large scale requires advanced wireless network architectures capable of accommodating all these novel functionalities. A particularly promising approach is cell-free massive MIMO (multiple-input multiple-output), or CF-mMIMO \cite{ngo2015cell, demir2021foundations}, which replaces traditional cell-based networks with a distributed system of cooperating access points (APs). These moderately-sized APs, linked to central processing units (CPUs), improve coverage, performance uniformity, and mobility support while minimizing interference. In user-centric CF-mMIMO \cite{buzzi2017cell, buzzi2017user}, each user equipment (UE) is served by a subset of APs, balancing macro-diversity, energy efficiency, and system complexity. Combining ISAC with CF-mMIMO enables distributed antenna systems for target sensing, where selected APs handle echo reception. This eliminates the need for full-duplex base stations, addressing a key issue in multi-cell mMIMO ISAC deployments \cite{buzzi2019using, liao2024powerallocation}.

This work explores the potential capabilities of CF-mMIMO in facilitating ISAC, concentrating on the system's scalability and the processes necessary for detecting and later tracking multiple targets within an operational ISAC wireless network. The current paper is an extended version of the earlier study \cite{buzzi2024scalable}, which presented initial results on the topic of scalability in CF-mMIMO systems performing ISAC. 

\subsection{State of the art} \label{sec:1.1}
ISAC with CF-mMIMO is a rapidly evolving research area, with recent studies exploring theoretical limits and practical aspects like power control, beamforming, and AP clustering. Despite notable progress, however, many challenges remain open \cite{lu2024isac, meng2024cooperative}.
In \cite{behdad2022power}, the authors present a CF-mMIMO network focused on downlink (DL) communication and occasional surveillance via a directed beam. They propose a power control scheme that maximizes sensing signal-to-noise ratio (SNR) while maintaining a minimum data rate. Results show their algorithm outperforms a communication-centric benchmark, underscoring the advantage of allocating power based on communication and sensing needs, instead of solely on communication. 
The work in \cite{chu2023integrated} considers a similar setting, and develops a complete signal model for an ISAC-enabled CF-mMIMO system using orthogonal frequency-division multiplexing (OFDM) modulation. It details the transceiver signal processing for both communication and radar functionalities, deriving expressions for UE signal-to-interference-plus-noise ratio (SINR) and sensing SNR. 
In \cite{babu2024precoding}, the authors focus on robust precoder design for ISAC in a multi-cell multi-user environment. The Cram\'er-Rao bound is used as a performance metric for sensing, and expressions for both coordinated beamforming and coordinated multipoint scenarios are derived.  
In \cite{demirhan2024cellfree}, the focus shifts towards the design of joint sensing and communication beamforming vectors in CF-mMIMO systems; the authors introduce communication-prioritized and sensing-prioritized beamforming as baseline approaches. They propose a joint beamforming strategy that seeks to balance these two tasks by maximizing the minimum per-user spectral efficiency (SE) alongside the sensing SNR. 
The problem of beamforming design is also addressed in \cite{mao2023beamforming}, by considering a beampattern optimization problem for the sensing task subject to data rate constraints.
Although these early studies employ basic models and assumptions, \cite{behdad2024isac} delves into target detection in CF-mMIMO systems, explicitly considering clutter interference and imperfect channel state information (CSI). It examines setups involving the use of communication beams for sensing and dedicated sensing beams, demonstrating that the integrated scheme outperforms orthogonal resource sharing.
The research in \cite{elfiatoure2023cell}, instead, addresses the problem of allocating the APs to either the sensing or the communication tasks; a joint power control and AP mode selection is thus considered, aimed at maximizing the minimum rate across UEs, subject to a prescribed minimum mainlobe-to-average-sidelobe ratio (MASR) level for sensing. 
An extended version of \cite{elfiatoure2023cell} is represented by \cite{elfiatoure2024multipletarget}, where, similarly to \cite{buzzi2024scalable}, the presence of multiple targets is accounted for, and a joint operation mode selection and power control design problem is formulated to maximize the SE fairness among the UEs while ensuring specific levels of MASR for sensing zones.
Moving towards a more theoretical perspective, the paper \cite{mao2024csregion} derives the communication-sensing (C-S) region for CF-mMIMO ISAC systems. The authors consider the impact of target location uncertainty and provide a closed-form derivation of the main statistics of the MIMO channel estimation error. Their findings reveal that the C-S region can be dynamically adjusted via the number of APs deployed, thus allowing for a flexible trade-off between communication and sensing within the network.

\subsection{Contributions} \label{sec:1.2}
Most of the referenced papers (with \cite{elfiatoure2024multipletarget} being the sole exception) focus on scenarios where a single target, located at a particular position, exists within the deployment zone of the CF-mMIMO system. These papers neither tackle the issue of \textit{scalability} when multiple targets are present or when a large area must be monitored. Scalability refers to the attribute that prevents the system's complexity and cost per dimension from increasing disproportionately as the network's size becomes infinitely large, which is essential for any workable system. While scalability for purely communication-based CF-mMIMO systems is well covered in \cite{interdonato2019scalability, bjornson2020scalable}, it has not, to our knowledge, been formulated for CF-mMIMO ISAC systems. Additionally, current research mainly addresses system design by focusing on performance criteria like the power focused on the target or related metrics such as MASR tied to beam pattern optimization. The design of detection algorithms and the impact of practical issues, like clutter, on the detector structure are infrequently discussed. Moreover, current research predominantly addresses sensing and detection tasks, often neglecting the aspect of tracking for targets that have already been identified. Tracking presents unique challenges, as targets may be situated anywhere, and their monitoring efforts can be severely hindered by strong interference from other targets.

The goal of this work is to address the above challenges and open problems. More precisely, the major contributions can be summarized as follows. The paper considers a CF-mMIMO system where a set of distributed APs perform ISAC tasks, detecting/tracking targets and communicating with UEs. The paper thus develops a unified framework and signal model for detecting potential targets and tracking the already detected targets, which may be in arbitrary positions. Novel detection and tracking signal processing algorithms are proposed by adopting the generalized likelihood ratio test (GLRT), which copes with the unknown response of the targets and with the unwanted target disturbance. Scalable association rules between APs and UEs/targets are presented, and the situation of simultaneous sensing for potential targets in several sensing zones is explicitly considered. In addition, the paper derives an innovative scalable power control mechanism that extends the fractional power control principles, traditionally applied to communication-only scenarios, to the ISAC context. This new approach allows for the fine-tuning of power allocation between communication functions and detection/tracking tasks, hence enabling a balance between the power dedicated to each UE and potential target. For benchmarking purposes, the paper also introduces a non-scalable power control optimization problem that seeks to maximize the minimum data rate across UEs while adhering to a quality of service (QoS) constraint for sensing performance. This non-scalable optimization problem is nonconvex and is addressed using the successive convex approximation (SCA) technique. The study is then completed with extensive numerical experiments, which evaluate the effectiveness of the proposed design concerning communication, detection, and tracking performance metrics. Simulation results demonstrate the satisfactory performance of the proposed detection and tracking algorithms. Besides, these outcomes also permit us to unveil the crucial role that other targets' interference plays in determining the system performance and to reveal existing trade-offs between communication and sensing in diverse scenarios.  

\subsection{Organization} \label{sec:1.3}
The remainder of this work is organized as follows. Section~\ref{sec:2} introduces the reference scenario and the scalability property for ISAC-enabled CF-mMIMO. Section~\ref{sec:3} describes the system model, while Sections~\ref{sec:4} and \ref{sec:5} derive the achievable data rate and transceiver signal processing for the ISAC functionalities, respectively. Section~\ref{sec:6} presents the power control schemes. Numerical results are provided in Section~\ref{sec:7} and, finally, concluding remarks are given in Section~\ref{sec:8}. 

\subsection{Notation} \label{sec:1.4}
In this work, scalars are denoted by italic letters. Boldface lowercase and uppercase letters denote vectors and matrices, respectively. The zero vector of length $m$ is denoted by $\bzero_m$. The identity matrix of size $m \times m$ is denoted by $\bI_m$. $\mathbb{C}^{m \times n}$ denotes the $m$ by $n$ dimensional complex space. The transpose, Hermitian, inverse, and trace operators are denoted by $(\cdot)^{\rm T}$, $(\cdot)^{\rm H}$, $(\cdot)^{{\rm -1}}$, and ${\rm tr}(\cdot)$, respectively. The Kronecker product is denoted by $\otimes$. The expectation operator is denoted by $\bbE[\cdot]$. The real and complex proper Gaussian distributions are denoted by $\calN(\cdot,\cdot)$ and $\calCN(\cdot,\cdot)$, respectively.

\section{Scalable CF-mMIMO for ISAC tasks} \label{sec:2}
We consider a distributed setting where a large number of APs are deployed to perform simultaneous communication with UEs together with surveillance of the surrounding environment in the same time-frequency slot. In this context, surveillance is indeed defined as the dual activities of detecting and tracking targets. Specifically, detection involves scanning the surroundings to identify new targets that have not been observed before. On the other hand, tracking pertains to focusing on previously identified targets to verify or update their predicted location and parameters.

We denote by $M$ the total number of APs, $K$ the number of UEs, and $L$ the number of targets. APs are equipped with $N$ antennas, while, for simplicity, UEs are assumed to have single-antenna transceivers. The APs are connected through a high-capacity fronthaul link to a CPU, where the centralized system operations occur. For simplicity, it is assumed all APs are connected to the same CPU; however, the case of multiple CPUs can be addressed as, for instance, in \cite{bjornson2020scalable}.

We concentrate on \textit{user-centric} CF-mMIMO systems \cite{buzzi2017cell, buzzi2017user, buzzi2019user}, or also scalable systems \cite{interdonato2019scalability, bjornson2020scalable}, where each UE is served only by a limited number of surrounding APs. According to \cite{bjornson2020scalable}, a CF-mMIMO system is scalable when the computational complexity required by channel estimation and transceiver signal processing at each AP, the required rate on each fronthaul link, and the computational complexity required at each AP/CPU to perform power allocation, stay finite even in the limiting case that the network size (i.e., the number of APs and the number of UEs) grows unboundedly.  
	
For an ISAC-enabled CF-mMIMO system, where both the sensing (detection/tracking) and communication tasks are to be accounted for, the above scalability definition is to be modified to account also for the system complexity generated by the sensing operations. We thus formulate the following

\noindent
\textbf{Definition:} \textit{An ISAC-enabled CF-mMIMO network is scalable if and only if the rate needed on each fronthaul link and the complexity required by the channel estimation and transceiver signal processing at each AP as well as the power allocation performed at each AP/CPU, stay finite even in the limiting case that the number of APs, the number of UEs, and the number of simultaneously detected/tracked targets grows unboundedly.}

\begin{figure}[t]
    \centerline{\includegraphics[scale=0.36]{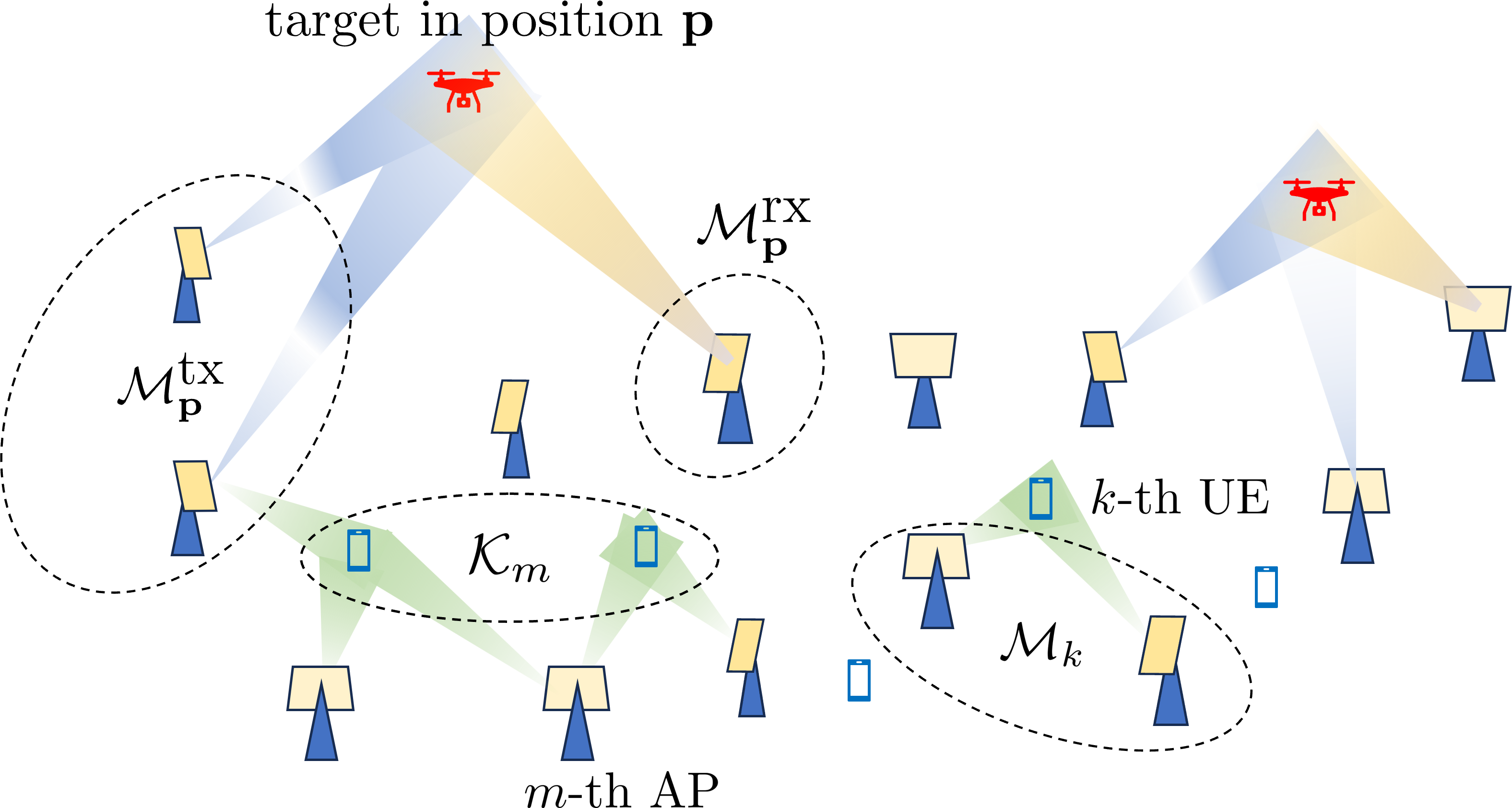}}    
    \caption{Illustrative setup. Several APs cooperate to perform communication and sensing tasks. The sets $\calM_{\bp}^{\rm tx}$ and $\calM_{\bp}^{\rm rx}$ refer to the transmit and receive APs inspecting position $\bp$, respectively. The set $\calK_m$ of UEs is served by AP $m$ and the set $\calM_k$ of APs is serving UE $k$.}
    \label{fig:PictorialRepresentation}
    \vspace{-1mm}
\end{figure}

In the following, to facilitate ISAC operation without relying on full-duplex capabilities, we will adopt a multi-static sensing approach \cite{chu2023integrated,behdad2024isac}, and assign specific APs exclusively for receiver sensing tasks. The APs are then classified into two categories: “transmit APs” and “receive APs”. Transmit APs are responsible for communication tasks like uplink (UL) training, UL data reception, DL data transmission, and sensing beams transmission. In turn, receive APs gather echoes from potential targets and contribute to UL data detection. Accordingly, we denote the set of transmit APs as $\cal{M}^{\rm tx}$, and the set of receive APs as $\calM^{\rm rx}$. Hence, the overall set of APs, $\calM = \calM^{\rm tx} \cup \calM^{\rm rx}$, has cardinality $M$.

\subsection{Scalable user-centric communication}
To guarantee scalability in communication, we employ a user-centric approach \cite{buzzi2017cell}, where each UE is served by a subset of APs. Let $\calM_k \subset \calM^{\rm tx}$ represent the set of APs serving UE $k$, and let $\calK_m$ denote the set of UEs served by AP $m$.

\subsection{Scalable target-centric sensing}
As discussed, we are interested in the case in which, apart from transmitting data, the system can simultaneously sense the presence of possible targets in several locations of the area to be surveilled (detection phase), and later monitor the multiple already detected targets (tracking phase). 

To ensure scalability, a \textit{target-centric} approach is required. Similar to communication, only some APs are responsible for detecting/tracking each target. Let $\bp$ be the spatial coordinates of the radar (range) cell to be inspected for detection/tracking purposes. We then denote by $\calM_{\bp}$ the set of APs involved in the detection/tracking task. More precisely, $\calM_{\bp}^{\rm tx} \triangleq \calM_{\bp}\cap \calM^{\rm tx}$ and $\calM_{\bp}^{\rm rx} \triangleq \calM_{\bp}\cap \calM^{\rm rx}$ are the sets containing the transmit and receive APs inspecting position $\bp$, respectively. See Fig.~\ref{fig:PictorialRepresentation} for a pictorial representation of the considered scenario.

For simplicity, we assume detection (discovery of potential new targets) and tracking (trajectory update of already detected targets) are performed in disjoint time intervals. The consideration of simultaneous detection and tracking can be investigated with a suitable generalization of the content of this paper and is left for future work. 

During the detection phase, the entire surveillance area is divided into regions (named as \textit{sensing regions}) of assigned size that do not overlap. At any given time, one radar range cell is inspected in each sensing region: otherwise stated, if we denote by $S$ the number of sensing regions, $S$ radar range cells are simultaneously inspected to discover new targets. This strategy permits inspecting the radar range cells in the entire surveillance area in an ordered way to avoid two close cells being inspected at the same time. We then denote by $\calS_m \subset \calS$ the subset of sensing regions inspected by transmit AP $m$, with $\calS$ the complete set of all clusters such that $\vert \calS \vert = S$. On the contrary, receive APs are responsible for only detecting a single target within their sensing region.

During the tracking phase, the goal is to confirm the presence of already detected targets and update their parameters. In this case, transmit AP $m$ tracks the detected targets within the subset $\calL_m$, and the receive AP focuses on one or more targets. Obviously, the network has no control over the position of the detected targets, and it can happen that close targets are to be tracked: mutual interference effects are thus to be properly taken into account. This is discussed further in Section~\ref{sec:4}.

\section{System Model} \label{sec:3}
We now describe the proposed scalable ISAC-enabled CF-mMIMO system model, focusing mainly on the sensing tasks: detection at positions $\bp_i$, with $i \in \calS$, and tracking at positions $\bp_l$, with $l \in \{1,\ldots,L\}$. Assuming block fading channel and OFDM modulation, each coherence block consists of $\tau_c$ time-frequency samples during which channels and reflections are constant and flat. This way, $t \in \{1,\ldots,\tau_c\}$ will denote the index for the symbols transmitted within a single coherence block to sense the positions $\bp_i$ (detection) or $\bp_l$ (tracking).

In that sense, $\bh_{k,m}$ ($\bh_{k,m}^{\rm H}$) denotes the $N$-dimensional UL (DL)\footnote{We assume that system operations happen within one channel coherence interval, and that time-division-duplex (TDD) protocol is used, to ensure equivalence (reciprocity) of UL and DL channels.} channel from UE $k$ to AP $m$; $\bG_{m,m'}$ is the $(N \times N)$-dimensional matrix representing the channel from AP $m' \in \calM^{\rm tx}$ to AP $m \in \calM^{\rm rx}$; and $\bH_{\bp,m,m'}$ refers to the $(N \times N)$-dimensional matrix for the composite channel linking transmit AP $m'$ to receive AP $m$ through the reflection from the target located in position $\bp$ (either $\bp_i$ or $\bp_l$). 

\subsection{Propagation Channels}\label{sec:2.1}
Under the above assumptions, the $N$-dimensional UL channel from UE $k$ to AP $m$ is \cite{interdonato2023coexistence}
\begin{equation}
    \bh_{k,m} \sim \calCN(\bzero,\bC_{k,m}),    
\end{equation}
where $\bC_{k,m} \in \mathbb{C}^{N \times N}$ refers to the spatial correlation matrix of the Rayleigh-distributed non-line-of-sight (NLoS) components. The corresponding large-scale fading coefficient (LSF), which includes path loss, is ${\rm tr}(\bC_{k,m})/N$.

For the AP-AP link, we adopt a Rician modeling:
\begin{equation}    
    \bG_{m,m'} = \sqrt{\frac{\ds b_{m,m'}}{\ds 1 + c_{m,m'} }}\left(\bar{\bG}_{m,m'} + \sqrt{c_{m,m'}} e^{j\psi_{m,m'}}\bV_{m,m'} \right),        
    \label{eq:Channel_AP} 
\end{equation}
where $b_{m,m'}$ is the LSF coefficient, $c_{m,m'}$ is the Rician factor, $\bar{\bG}_{m,m'} \in \mathbb{C}^{N \times N}$ contains the correlated\footnote{In line with \cite{behdad2022power}, a popular choice for the channel covariance matrix $\bQ_{m,m'}$ is the well-known Kronecker model. However, we leave this unspecified until the simulations section to keep our analysis general.} NLoS components such that $\bar{\bg}_{m,m'} = {\rm vec}(\bar{\bG}_{m,m'}) \sim \calCN(\bzero_{N^2},\bQ_{m,m'})$, $\psi_{m,m'}\sim \calU[0,2\pi]$ is the phase offset, and $\bV_{m,m'} \in \mathbb{C}^{N \times N}$ is the equivalent array response at the LoS direction:
\begin{equation}
    \bV_{m,m'} = \ba_m\left(\varphi_{m,m'},\theta_{m,m'}\right) \ba_{m'}^{\rm H}\left(\varphi_{m',m},\theta_{m',m}\right),    
\end{equation}
with $\ba_m(\varphi_{m,m'},\theta_{m,m'})$ the response of the array (or steering vector) for the azimuth $\varphi_{m,m'}$ ($\varphi_{m',m}$) and elevation $\theta_{m,m'}$ ($\theta_{m',m}$) angles of arrival (departure) from AP $m$ to AP $m'$.

According to \cite{dandrea2020uav}, the Rician factors can be defined as
\begin{equation}
    c_{m,m'} = \frac{\ds p_{\rm LoS}\left(d_{m,m'}\right)}{\ds 1 - p_{\rm LoS}\left(d_{m,m'}\right)},    
\end{equation}
where $p_{\rm LoS}(d_{m,m'})$ is the LoS probability that depends on the distance $d_{m,m'}$ between AP $m$ and $m'$ \cite[Table B.1.2.1-2]{3GPP36814}.

Finally, assuming LoS propagation between the targets and the involved APs, a convenient expression for $\bH_{\bp,m,m'}$ is
\begin{equation}
    \bH_{\bp,m,m'}= \tilde{\alpha}_{\bp,m,m'} \underbrace{\ba_m\left(\varphi_{m,\bp},\theta_{m,\bp}\right) 
    \ba_{m'}^{\rm H}\left(\varphi_{m',\bp},\theta_{m',\bp}\right)}_{\bA_{\bp,m,m'}},
    \label{eq:Channel_target}
\end{equation}
where $\tilde{\alpha}_{\bp,m,m'}=\alpha_{\bp,m,m'} \sqrt{\beta_{\bp,m,m'}}$ is a complex scalar coefficient, with $\alpha_{\bp,m,m'}$ the target reflectivity, or radar cross-section (RCS), and $\beta_{\bp,m,m'}$ the product of the path loss from transmit AP $m'$ to the target at position $\bp$ and that from the target to receive AP $m$. As previously mentioned, we follow the Swerling-I model for the RCS, in which $\alpha_{\bp,m,m'}$ is kept constant over consecutive symbols within the coherence block \cite{behdad2022power}. Moreover,  $\varphi_{m,\bp},\theta_{m,\bp}$ are the azimuth and elevation angle of the position $\bp$ with respect to (w.r.t.) the antenna array of AP $m$. A similar meaning have the quantities $\varphi_{m',\bp}$ and $\theta_{m',\bp}$ w.r.t. AP $m'$.

\subsection{Uplink Channel Estimation}\label{sec:2.2}
Assuming perfect CSI can be overly optimistic in many applications. In practice, obtaining such channel knowledge locally at the APs via UL orthogonal pilots is more realistic. This approach enables the characterization of the sufficient statistics of the channels \cite{ngo2017cell}.

A feasible option could be the minimum mean-square error (MMSE) estimation \cite[Subsection~V-B]{liesegang2024emf}:
\begin{equation}
    \hat{\bh}_{k,m} = \frac{1}{\ds \sqrt{\tau_p \bar{\eta}_k}}\bm{\Lambda}_{k,m} \bm{\varphi}_{k,m},    
\end{equation}
where $\bm{\Lambda}_{k,m} = \tau_p \bar{\eta}_k \bC_{k,m} \bm{\Gamma}_{k,m}^{-1}$ is the MMSE matrix, $\tau_p$ is the number of pilots, $\bar{\eta}_k$ is the UL training power, and
\begin{equation}
    \bm{\Gamma}_{k,m} = \sum\nolimits_{j = 1}^K \tau_p \bar{\eta}_j \bC_{j,m} \left\vert \bm{\pi}_j^{\rm H} \bm{\pi}_k \right\vert^2 + \sigma_m^2 \bI_L,            
\end{equation}
denotes the covariance matrix of the least-squares observations $\bm{\varphi}_{k,m}  \in \mathbb{C}^N$ providing sufficient statistics, i.e.,
\begin{equation}    
    \bm{\varphi}_{k,m} = \sum\nolimits_{j = 1}^K \sqrt{\tau_p \bar{\eta}_j} \bh_{j,m} \bm{\pi}_j^{\rm H} \bm{\pi}_k + \bm{\omega}_{k,m} ,        
\end{equation}
with $\bm{\pi}_k \in \mathbb{C}^{\tau_p}$ the sequence of pilots sent by UE $k$ to estimate the channel, and $\bm{\omega}_{k,m} \in \mathbb{C}^N$ the equalized ambient noise with variance $\sigma_{\omega}^2$ \cite[Subsection~II-B]{interdonato2023coexistence}. 

\subsection{Downlink Transmission} \label{sec:3.1}
Assuming to be in the DL transmission and sensing phase, AP $m \in \calM^{\rm tx}$ sends data to the UEs in the set $\calK_m$ plus additional beams to first, detect the presence of potential targets within the surveillance area and later, track their location. 

The baseband equivalent of the $t$-th transmit signal with power $P_m$ can be represented as
\begin{equation}
    \bs_m[t] = \sum\nolimits_{k \in \calK_m}\sqrt{\eta_{k,m}} \bw_{k,m} x_k [t] + \bs_{0,m}[t],
    \label{eq:s_m}
\end{equation}
where $\eta_{k,m}$, $\bw_{k,m}$, and $x_k[t]$ are the power, the unit precoder, and the unit-energy complex number representing the data symbol intended for UE $k$, respectively. 

Let $P_m^{\rm c}$ denote the portion of power dedicated to communication, i.e., $\sum_{k \in \calK_m} \eta_{k,m} = P_m^{\rm c}$. The independent sensing (probing) signal $\bs_{0,m}$ initially used for detection and afterward for tracking will have power $P_m^{\rm s}$ such that $P_m^{\rm c} + P_m^{\rm s} \leq P_m$. The expression of $\bs_{0,m}$ will differ depending on the task.

\subsubsection*{Detection}
When focusing the transmit beams at positions $\bp_i$ within the sensing regions set $\calS_m$, we have
\begin{equation}
    \bs_{0,m}[t] = \sum\nolimits_{i \in \calS_m} \sqrt{\mu_{i,m}} \bw_{0,m}(\bp_i) x_{i,m}[t],    
\end{equation}
where $\mu_{i,m}$ are the powers used for the detection such that $\sum_{i \in \calS_m} \mu_{i,m} = P_m^{\rm s}$, $\bw_{0,m}(\bp_i)$ is the beamforming unit vector used to sense the potential target in position $\bp_i$, and $x_{i,m}[t]$ is a fake unit-norm data symbol associated to the $i$-th beam.

\subsubsection*{Tracking}
Once the position of the target is estimated, the system uses the additional signal for tracking, i.e.,
\begin{equation}
    \bs_{0,m}[t] = \sum\nolimits_{l \in \calL_m} \sqrt{\mu_{l,m}} \bw_{0,m}(\hat{\bp}_l) x_{l,m}[t], 
    \label{eq:s_0m}
\end{equation}
where $\mu_{l,m}$, $\bw_{0,m}(\hat{\bp}_l)$, and $x_{l,m}[t]$ are defined likewise. In contrast, $\hat{\bp}_l$ is now the estimated location of the $l$-th target. 

Working with Cartesian coordinates leads to coupled errors that are difficult to characterize \cite[Section~18.10]{richards2023principles}. For that reason, since we obtain $\hat{\bp}_l$ from the angular location, we will concentrate on the following estimate: 
\begin{equation}
    \left[\varphi_{m,\hat{\bp}_l} \theta_{m,\hat{\bp}_l}\right] = \left[\varphi_{m,\bp_l} \theta_{m,\bp_l}\right] + \bm{\varepsilon}_l,    
    \label{eq:Position_tracking}
\end{equation}
with $\bm{\varepsilon}_l$ the angle estimation error that is sometimes modeled as an uncorrelated Gaussian vector with zero mean and (known) covariance matrix $\bm{\Delta}_l$, i.e., $\bm{\varepsilon}_l \sim \calN(\bzero,\bm{\Delta}_l)$ \cite[Section~18.9]{richards2023principles}. However, other statistics might be used for $\bm{\varepsilon}_l$ depending on the angle estimation strategy (which is beyond the scope of this work). Therefore, we will consider a general probability density function (PDF) $f(\bm{\varepsilon}_l)$ for the upcoming analysis and continue the discussion in Section~\ref{sec:7}. 

\subsection{Beamformers Definition}\label{sec:4.1}
In this subsection, we discuss the beamformers used for communication, detection, and tracking tasks.
For the DL data transmission, we assume maximum ratio transmission (MRT):
\begin{equation}
    \bw_{k,m} = \hat{\bh}_{k,m}/\sqrt{\bbE[\|\hat{\bh}_{k,m}\|^2]}.
    \label{eq:MRT_BF}
\end{equation}

For the sensing (both detection and tracking), since we have perfect knowledge of the position to be inspected, we assume the APs point the probing signal towards the angles $\varphi_{m,\bp_i}$ and $\theta_{m,\bp_i}$ for detection and $\varphi_{m,\hat{\bp}_l}$ and $\theta_{m,\hat{\bp}_l}$ for sensing. That is,
\begin{equation}
    \bw_{0,m}(\bp_i) = \ba_m(\varphi_{m,\bp_i},\theta_{m,\bp_i}),
    \label{eq:detection_BF}
\end{equation}
while $\bw_{0,m}(\hat{\bp}_l)$ is obtained by interchanging index $i$ by $l$ and position $\bp_i$ by $\hat{\bp}_l$ in \eqref{eq:detection_BF}. Recall that in all cases, one can select other processing alternatives, e.g., zero-forcing as in \cite{behdad2024isac,buzzi2024scalable}. In this work, we advocate for the previous scalable and low-complexity beamforming at the APs, considering the volume of all tasks they implement.

\section{Achievable Data Rate} \label{sec:4}
Regarding the communication task, UE $k$ receives the following scalar observable:
\begin{equation}
    y_k[t] = \sum\nolimits_{m \in \calM^{\rm tx}}\bh_{k,m}^{\rm H} \bs_m[t] + z_k[t],
    \label{eq:y_k}
\end{equation}
with $z_k[t] \sim \calCN(0, \sigma^2_z)$ the additive noise contribution. 

For imperfect CSI, a tractable rate expression can be obtained via the \textit{use-and-then-forget} bound, widely used in the mMIMO literature. Essentially, it implies channel estimates are only exploited for the computation of the beamformers and then dumped during signal detection (cf. \cite[(11)]{interdonato2023coexistence}).

Concisely, substituting \eqref{eq:s_m} in \eqref{eq:y_k} and assuming the CPU has channel distribution information (CDI) and no knowledge of the realizations, the received signal of UE $k$ yields
\begin{equation}
    \begin{aligned}
        y_k[t] &= \bar{\calH}_{k} x_k[t] + \left(\tilde{\calH}_{k} - \bar{\calH}_{k} \right) x_k[t] + z_k[t] \\ &\quad + 
        \sum\nolimits_{j\neq k} \sum\nolimits_{m \in \calM_j} \sqrt{\eta_{j,m}} \bh_{k,m}^{\rm H}\bw_{j,m} x_j[t] \\
        &\quad +
        \sum\nolimits_{m \in \calM^{\rm tx}} \bh_{k,m}^{\rm H} \bs_{0,m}[t],
    \end{aligned}    
    \label{eq:y_k_UatF}
\end{equation}    
with
\begin{equation*}
    \begin{aligned}
        \bar{\calH}_{k} &= \bbE\left[\sum\nolimits_{m \in \calM_k} \sqrt{\eta_{k,m}} \bh_{k,m}^{\rm H}\bw_{k,m}\right]\\
        \tilde{\calH}_{k} &= \sum\nolimits_{m \in \calM_k} \sqrt{\eta_{k,m}} \bh_{k,m}^{\rm H}\bw_{k,m}.
    \end{aligned}    
\end{equation*}

In \eqref{eq:y_k_UatF}, the first three terms represent the useful, uncertainty, and noise signals, respectively, while the second and third lines represent the interference coming from communication signals intended for other UEs and the sensing signals, respectively. In writing \eqref{eq:y_k_UatF}, we have neglected the signal component reflected by potential targets in the area, since it is reasonable to assume they are extremely weak and can be included in the multipath contributing to the formation of the channels $\bh_{k,m}$.

Following the rationale in \cite[Corollary~6.3]{demir2021foundations}, we can treat the estimation errors as additional noise and consider the worst-case scenario (uncorrelated Gaussian model). This way, under the assumption of standard normal codebooks ($x_k[t],x_{i,m}[t] \sim \calCN(0,1)$ $\forall t$), the lower bound for the data rate results: 
\begin{equation}
    R_k = \frac{\ds \tau_d}{\ds \tau_c} B \log_2\left(1 + \gamma_k\right),    
\end{equation}
with $\tau_d$, $\tau_c$, and $B$ the number of DL transmit symbols, the number of (time-frequency) samples per coherence block, and the system's bandwidth, respectively. The communication SINR $\gamma_k$ is given by
\begin{equation}    
    \gamma_k = \frac{\calA_k}{\calB_k + \calC_k + \calD_k + \sigma_z^2},           
    \label{eq:commSINR}
\end{equation}
where, considering the independence among transmit signals (in terms of data and sensing components), we have
\begin{equation*}
    \begin{aligned}        
        \calA_k &\triangleq \bar{\calH}_{k}^2, \quad   \calB_k \triangleq \bbE\left[\left\vert \tilde{\calH}_{k}\right\vert^2 \right] - \calA_k, \\
        \calC_k &\triangleq \sum\nolimits_{j \neq k} \bbE\left[\left\vert \sum\nolimits_{m \in \calM_j} \sqrt{\eta_{j,m}} \bh_{k,m}^{\rm H}\bw_{j,m} \right\vert^2 \right], \\
        \calD_k &\triangleq \sum\nolimits_{m \in \calM^{\rm tx}} \sum\nolimits_{i \in  \calS_m} \mu_{i,m} \bbE\left[ \bh_{k,m}^{\rm H} \bw_{0,m}(\bp_i)\right]^2.       
    \end{aligned}    
\end{equation*}

In general, closed-form expressions for the moments within \eqref{eq:commSINR} are difficult to find. However, following the arguments in \cite[Corollary~6.4]{demir2021foundations}, we can derive the moments under the assumption of MRT in \eqref{eq:MRT_BF}. In particular, provided the sole Rayleigh distribution of the NLoS channels, we can exploit the relationship between jointly Gaussian vectors \cite[Lemma~B.5]{demir2021foundations}
\begin{equation}
    \bbE\left[\left\vert \bx^{\rm H}\by \right\vert^2\right] = \rm{tr}\left(\bbE \left[\bx \bx^{\rm H}\right]\bbE \left[\by \by^{\rm H}\right]\right) + \left\vert \rm{tr}\left(\bbE \left[\by \bx^{\rm H}\right]\right)\right\vert^2,
    \label{Gaussian_vectors_rel}
\end{equation}
for any circularly-symmetric complex normals $\bx$ and $\by$.

By considering \eqref{Gaussian_vectors_rel} and the modeling from Section~\ref{sec:3}, after some manipulations, one can easily compute $\calA_k$, $\calB_k$, and $\calC_k$. Regarding the derivation of the sensing interference $\calD_k$, note that this additional interference does not appear in typical CF-mMIMO (or cellular mMIMO) deployments but only in ISAC-based systems; thus, it is exclusive to our scenario. The sensing contribution $\bW_{i,m}$ ($\bW_{l,m}$), independent of the UE channels, is derived below for the detection (tracking) beamformers. 
\subsubsection{Detection} This sensing phase leads to the following:
\begin{equation}    
    \bW_{i,m} = \bbE_{\calP_i}\left[\bw_{0,m}(\bp_i)\bw_{0,m}(\bp_i)^{\rm H}\right],    
    \label{eq:W_detection}
\end{equation}
where the expectation is taken over the long-term distribution $\calP_i$ of the inspected locations $\bp_i$ along the data communication. This can be calculated via Monte Carlo simulations.
\subsubsection{Tracking} Changing index $i$ by $l$ and position $\bp_i$ by $\hat{\bp}_l$,
\begin{equation}
    \bW_{l,m} = \bbE_{\calP_l}\left[\bbE_{\calE_l}\left[\bw_{0,m}(\hat{\bp}_l)\bw_{0,m}(\hat{\bp}_l)^{\rm H} \vert \bp_l \right]\right],
    \label{eq:W_tracking}
\end{equation}
where, per each target spatial distribution, instead of a constant (known) direction, we also need to average over the statistics $\calE_l$ of the estimation errors $\bm{\varepsilon}_l$. For instance, for a uniform linear array with a $d$ element spacing \cite{demir2021foundations}, the elements of the inner expectation $\bbE_{\calE_l}[\bw_{0,m}(\bp_l)\bw_{0,m}(\bp_l)^{\rm H}]_{u,u'}$ yield (cf. \eqref{eq:Position_tracking})
\begin{equation}    
        \int e^{j 2 \pi \left(u - u'\right) d \sin\left(\varphi_{m,\hat{\bp}_l}\right)\cos\left(\theta_{m,\hat{\bp}_l}\right)} f\left(\bm{\varepsilon}_l\right) d\bm{\varepsilon}_l.
\end{equation}

As mentioned earlier, the PDF $f(\bm{\varepsilon}_l)$ depends on each angle estimation technique (which is left for future studies). For the time being, we will then assume \eqref{eq:W_tracking} is also computed offline numerically. A more comprehensive debate is provided later.

The complete expression of the communication SINR $\gamma_k$ for UE $k$ assuming MRT at the APs for data communication is reported in \eqref{eq:gamma_k} at the top of the next page for the tracking phase (the detection case can be obtained analogously), where $\bm{\Phi}_{k,m} = \bm{\Lambda}_{k,m}\bC_{k,m}$ and $\bW_{l,m}$ is expressed according to \eqref{eq:W_tracking}.

\begin{figure*}[t]
    \begin{equation}
        \gamma_k = \frac{\ds \left\vert \sum\nolimits_{m \in \calM_k} \sqrt{\eta_{k,m} {\rm tr}\left(\bm{\Phi}_{k,m}\right)} \right\vert^2}{\ds \sum_{j = 1}^K\sum_{m \in \calM_j} \eta_{j,m} \frac{\ds {\rm tr}\left(\bC_{k,m} \bm{\Phi}_{j,m}\right)}{\ds {\rm tr}(\bm{\Phi}_{j,m})} + \sum_{j \neq k} \left\vert \bm{\pi}_j^{\rm H} \bm{\pi}_k  \sum_{m \in \calM_j} \sqrt{\eta_{j,m}} \frac{\ds {\rm tr}\left(\bC_{k,m} \bm{\Lambda}_{j,m}\right)}{\ds \sqrt{{\rm tr}(\bm{\Phi}_{j,m})}} \right\vert^2 + \sum_{m \in \calM^{\rm tx}} \sum_{l \in \calL_m} \mu_{l,m} {\rm tr}\left( \bC_{k,m} \bW_{l,m} \right) +  \sigma^2_z},
        \label{eq:gamma_k}
    \end{equation} 
    \hrule    
\end{figure*}

\section{GLRT-based Processing} \label{sec:5}
\subsection{Detection}
Let us now consider the detection task and let us denote by $\bar{\by}_m$ the $N$-dimensional signal received at AP $m \in \calM_{\bp_i}^{\rm rx} = \calM_{\bp_i} \cap \calM^{\rm rx}$ to detect the possible presence of a target at position $\bp_i$ in the $i$-th detection region. We introduce a binary random variate, $a(\bp_i) \in \{0, 1\}$, which equals 1 if a target is present at position $\bp_i$, and 0 otherwise. Accordingly, we have:
\begin{equation}
    \begin{aligned}
        \bar{\by}_m[t] &= \sum\nolimits_{i =1}^S a(\bp_i) 
        \sum\nolimits_{m' \in \calM^{\rm tx}} \bH_{i,m,m'} \bs_{m'}[t]  \\ 
        &\quad + \sum\nolimits_{m' \in \calM^{\rm tx}} \bG_{m,m'} \bs_{m'}[t] + \tilde{\bz}_m[t],
    \end{aligned}
    \label{eq:y_bar_m}
\end{equation}
with $\bH_{i,m,m'} \equiv \bH_{\bp_i,m,m'}$ for ease of notation. $\tilde{\bz}_m[t]\sim \calCN(\bzero, \sigma^2_z \bI_N)$ is the additive thermal noise, whereas the term $\sum\nolimits_{m' \in \calM^{\rm tx}} \bG_{m,m'} \bs_{m'}$ represents the direct signals that propagate from the transmit APs in the set $\calM^{\rm tx}$ to receive AP $m$ responsible for the detection. Given that all APs in the system are controlled by the network (in this case, they are even linked to the same CPU), it is reasonable to assume that the receiving AP knows the transmitted signal. Motivated by \cite{behdad2024isac}, we also consider that only the LoS components of the channels $\bG_{m,m'}$ in \eqref{eq:Channel_AP}, for all $(m,m') \in \calM^{\rm rx} \times \calM^{\rm tx}$ have been perfectly estimated and known to the system. 

Under these hypotheses, and after perfectly subtracting these terms from $\bar{\by}_m$, we end up with the observable
\begin{equation}
    \begin{aligned}
        \tilde{\by}_m[t] &= \sum\nolimits_{i =1}^S a(\bp_i) \sum\nolimits_{m' \in \calM^{\rm tx}} \bH_{i,m,m'} \bs_{m'}[t] \\ &\quad + \sum\nolimits_{m' \in \calM^{\rm tx}} \varkappa_{m,m'} \bar{\bG}_{m,m'} \bs_{m'}[t] + \tilde{\bz}_m[t],
    \end{aligned}
    \label{eq:y_tilde_m}
\end{equation}
with $\varkappa_{m,m'} = \sqrt{b_{m,m'}/(1 + c_{m,m'})}$ for brevity. In \eqref{eq:y_tilde_m}, the unknown clutter $\bar{\bG}_{m,m'}$ (NLoS paths) caused by temporary obstacles between APs can be seen as additional noise.

Now, based on the observations $\tilde{\by}_m$, for all $m \in \calM^{\rm rx}$, the network has to decide on the possible presence of targets in positions $\bp_1, \ldots, \bp_S$. Optimal execution of this task would require the collection of all the $N$-dimensional observables $\left\{\tilde{\by}_m, m \in \calM^{\rm rx}\right\}$ at the CPU and the execution of a detection test with $2^S$ hypotheses. This approach is, however, unscalable as the network size increases, so here we propose a different suboptimal, but scalable approach. 

First, instead of focusing on a joint detection test with $2^S$ hypotheses, we perform $S$ disjoint binary hypothesis tests. Let us focus on the $i$-th test for the target in position $\bp_i$. More precisely, we consider the observables $\tilde{\by}_m$ for all $m \in \calM_{\bp_i}^{\rm rx}$, and, for the current detector design stage, we neglect the much weaker echo from potential targets at (far) positions $\bp_{i' \neq i}$. The probability of other targets being simultaneously located in the same range cell decreases with the size of the deployment area. Essentially, this means every target is usually detected in separate time slots (i.e., without interference). 

For detecting the target at the inspected rangecell position $\bp_i$, we thus have the following signals:
\begin{equation}
    \begin{aligned}
        \tilde{\by}_m[t] &\approx a(\bp_i) \sum\nolimits_{m' \in \calM^{\rm tx}} \bH_{i,m,m'} \bs_{m'}[t] \\ &\quad + \sum\nolimits_{m' \in \calM^{\rm tx}} \varkappa_{m,m'} \bar{\bG}_{m,m'} \bs_{m'}[t] + \tilde{\bz}_m[t],
    \end{aligned}
    \label{eq:y_tilde_m_approx}
\end{equation}
for all $m \in \calM_{\bp_i}^{\rm rx}$. Recall that, given \eqref{eq:Channel_target}, the observable $\tilde{\by}_m$ contains the RCSs $\alpha_{i,m,m'} \equiv \alpha_{\bp_i,m,m'}$, for all $m' \in \calM^{\rm tx}$, which we model as unknown deterministic parameters.

Let us define the $N \times \left\vert\calM^{\rm tx}\right\vert$-dimensional matrix $\bD_{i,m}[t]$ containing on its columns the vectors
\begin{equation}
     \bd_{i,m,m'}[t] = \sqrt{\beta_{i,m,m'}} \bA_{i,m,m'} \bs_{m'}[t],    
    \label{eq:d_columns}
\end{equation}
with $m' \in \{1, \ldots, \left\vert\calM^{\rm tx}\right\vert\}$, $\beta_{i,m,m'} \equiv \beta_{\bp_i,m,m'}$, $\bA_{i,m,m'} \equiv \bA_{\bp_i,m,m'}$, and the $\left\vert\calM^{\rm tx}\right\vert$-dimensional vector
\begin{equation}
     \bm{\alpha}_{i,m}=\left[\alpha_{i,m,1},\ldots, \alpha_{i,m,\left\vert\calM^{\rm tx}\right\vert}\right]^{\rm T}.
\end{equation}

Similarly, for notational convenience, we denote
\begin{equation}
    \tilde{\bg}_m[t] = \sum\nolimits_{m' \in \calM^{\rm tx}}\varkappa_{m,m'}\left(\bs_{m'}^{\rm T}[t] \otimes \bI_N\right) \bar{\bg}_{m,m'},
\end{equation}
where we used the identity ${\rm vec}(\bX \bY \bZ) = (\bZ^{\rm T} \otimes \bX){\rm vec}(\bY)$ with $\bX = \bI_N$, $\bY = \bar{\bG}_{m,m'}$, and $\bZ = \bs_{m'}[t]$ \cite[(520)]{petersen2012matrix}.

Based on the above definitions, \eqref{eq:y_tilde_m_approx} can be written as
 \begin{equation}  
    \tilde{\by}_m[t] \approx a(\bp_i)  \bD_{i,m}[t] \bm{\alpha}_{i,m} + \tilde{\bg}_m[t] + \tilde{\bz}_m[t] \triangleq \tilde{\by}_{i,m}[t].    
\end{equation}

After collecting $\tau_s \leq \tau_c$ sensing samples, we thus formulate the detection problem for the radar cell centered at position $\bp_i$ at AP $m$ as the following binary test
\begin{equation}
    \left\{\begin{array}{ll}
        H_1: & \ddot{\by}_{i,m} = \ds \ddot{\bD}_{i,m} \bm{\alpha}_{i,m}  + \ddot{\bg}_m + \ddot{\bz}_m \\
        H_0: & \ddot{\by}_{i,m} = \ds \ddot{\bg}_m + \ddot{\bz}_m ,
    \end{array} \right.
    \label{eq:HT_detection}
\end{equation}
where $\ddot{\by}_{i,m} \in \mathbb{C}^{N \tau_s}$, $\ddot{\bD}_{i,m} \in \mathbb{C}^{N \tau_s \times \vert \calM^{\rm tx} \vert}$, $\ddot{\bg}_m \in \mathbb{C}^{N \tau_s}$, and $\ddot{\bz}_m \in \mathbb{C}^{N \tau_s}$ are the concatenation of $\tilde{\by}_{i,m}[t]$, $\tilde{\bD}_{i,m}[t]$, $\tilde{\bg}_m[t]$, and $\tilde{\bz}_m[t]$ for $t = \{1, \ldots, \tau_s\}$, respectively. 

To solve the binary hypothesis test in \eqref{eq:HT_detection}, we first notice that the observables in hypothesis $H_1$ depend on the unknown vector $\bm{\alpha}_{i,m}$, which must be properly accounted for. Two alternatives are possible: either $\bm{\alpha}_{i,m}$ is modeled as a correlated complex Gaussian random vector (cf. \cite{behdad2024isac}), and this alters the PDF of the observables under the hypothesis $H_1$; or $\bm{\alpha}_{i,m}$ is modeled as an unknown, but deterministic, parameter. In this paper, to avoid having a detector structure tailored to a specific statistical model for the target reflectivity, we follow the latter approach\footnote{At the performance analysis stage, instead, $\bm{\alpha}_{i,m}$ will be modeled as a correlated complex Gaussian vector.} and, thus, solve the test in \eqref{eq:HT_detection} through the GLRT (generalized likelihood ratio test), which amounts to substituting the maximum-likelihood (ML) estimate of the vector $\bm{\alpha}_{i,m}$ in the likelihood ratio test and comparing the result with a suitable threshold $\delta_i$ \cite{li2008mimo}: 
\begin{equation}
    \frac{\ds {\rm max}_{\bm{\alpha}_{i,m}} \, f\left(\ddot{\by}_{i,m} \vert H_1, \bm{\alpha}_{i,m},\ddot{\bQ}_m\right)}{\ds f\left(\ddot{\by}_{i,m} \vert H_0, \ddot{\bQ}_m\right)} \test \delta_i,
    \label{eq:GLRT_initial}
\end{equation}
where $\ddot{\bQ}_m = \sum_{m' \in \calM^{\rm tx}} \varkappa_{m,m'}^2 (\ddot{\bS}_{m'} \otimes \bI_N)\bQ_{m,m'}(\ddot{\bS}_{m'}^{\rm H} \otimes \bI_N)$ is the covariance matrix of the aggregate clutter $\ddot{\bg}_m$, with $\ddot{\bS}_{m'} = [\bs_{m'}[1],\ldots, \bs_{m'}[\tau_s]]^{\rm T}$ the concatenation of (known) transmit signals from AP $m'$.

Since when conditioning on $\bm{\alpha}_{i,m}$ ($\ddot{\bQ}_m$ is indeed known), the observable has a complex Gaussian distribution, it is easy to show that the GLRT test can be written as in \eqref{eq:GLRT_detection} at the top of the next page, where $\Re \lbrace \cdot \rbrace$ denotes the real part operator and $\bm{\Psi}_m = \ddot{\bQ}_m + \sigma_z^2 \bI_{N \tau_s}$ is the covariance matrix of the aggregate clutter plus the original noise.

\begin{figure*}
    \begin{equation}    
        \min_{\bm{\alpha}_{i,m}}\left[\bm{\alpha}_{i,m}^{\rm H} \ddot{\bD}_{i,m}^{\rm H} \bm{\Psi}_m^{-1} \bD_{i,m} \bm{\alpha}_{i,m} - 2 \Re \left\lbrace \bm{\alpha}_{i,m}^{\rm H} \ddot{\bD}_{i,m}^{\rm H} \bm{\Psi}_m^{-1} \ddot{\by}_{i,m}\right\rbrace\right] \test \ln(\delta_i)    
        \label{eq:GLRT_detection}
    \end{equation}
    \hrule
\end{figure*}

The minimization in \eqref{eq:GLRT_detection} w.r.t. $\bm{\alpha}_{i,m}$ can be computed in closed-form and the minimum is obtained at
\begin{equation}
    \hat{\bm{\alpha}}_{i,m} = \left( \ddot{\bD}_{i,m}^{\rm H} \bm{\Psi}_m^{-1} \ddot{\bD}_{i,m}\right)^{-1} \ddot{\bD}_{i,m}^{\rm H} \bm{\Psi}_m^{-1} \ddot{\by}_{i,m}.
    \label{eq:alpha_opt_detection}
\end{equation}

Substituting \eqref{eq:alpha_opt_detection} in the test \eqref{eq:GLRT_detection}, and integrating at the CPU the contribution from all the APs in the set $\calM_{\bp_i}^{\rm rx}$, we obtain the final GLRT as
\begin{equation}
    \sum\nolimits_{m \in \calM_{\bp_i}^{\rm rx}} \norm{\bU_{i,m}^{\rm H}\bm{\Psi}_m^{-0.5}\ddot{\by}_{i,m}}^2 \test \ln \left(\delta_i\right) ,    
    \label{eq:GLRT_detection_CPU}
\end{equation}
with $\bU_{i,m} \in \mathbb{C}^{N \tau_s \times r_{i,m}}$ the semi-unitary matrix of the left-singular vectors from $\bm{\Psi}_m^{-0.5}\ddot{\bD}_{i,m}$ associated to the $r_{i,m} = {\rm rank}(\bm{\Psi}_m^{-0.5}\ddot{\bD}_{i,m}) \leq {\rm min}(N \tau_s,\vert \calM^{\rm tx} \vert)$ non-zero singular values. In CF deployments, assuming $N \leq \vert \calM^{\rm tx} \vert$ is reasonable since the number of serving APs typically exceeds the number of transmit antennas. Thus, for $\tau_s = 1$, the linear system of equations $H_1$ in \eqref{eq:HT_detection} is undetermined, leading to \eqref{eq:GLRT_detection} equal to zero \cite[Subsection~10.3.1]{li2008mimo}. As a result, no ML estimation can be performed, which is why including sensing samples is essential (a single time symbol does not suffice to retrieve all RCSs). From now on, we safely assume $N \tau_s \geq \vert \calM^{\rm tx} \vert$ is satisfied. Due to the definition in \eqref{eq:d_columns}, $\ddot{\bD}_{i,m}$ will be a full-column-rank matrix and $r_{i,m} \leq \vert \calM^{\rm tx} \vert$ (the joint covariance $\bm{\Psi}_m$ is also a full-rank matrix).

Given the test \eqref{eq:GLRT_detection_CPU}, we can define the receive sensing signal-to-clutter-plus-noise ratio (SCNR) for the range cell at position $\bp_i$ in the $i$-th sensing area, say $\gamma_{\bp_i} $, as the ratio between the power of the useful signal in $\bU_{i,m}^{\rm H}\bm{\Psi}_m^{-0.5}\ddot{\by}_{i,m}$ and the power of its clutter plus noise components, i.e.,
\begin{equation}
    \begin{aligned}
        \gamma_{\bp_i}  &= \frac{\ds \bbE \left[\sum\nolimits_{m \in \calM_{\bp_i}^{\rm rx}} \norm{\bU_{i,m}^{\rm H} \bm{\Psi}_m^{-0.5}\ddot{\bD}_{i,m} \bm{\alpha}_{i,m}}^2 \right]}
        {\ds \bbE \left[ \sum\nolimits_{m \in \calM_{\bp_i}^{\rm rx}} \norm{\bU_{i,m}^{\rm H} \bm{\Psi}_m^{-0.5} \left(\ddot{\bg}_m + \ddot{\bz}_m\right)}^2 \right]} \\
        &= \frac{\ds \sum\nolimits_{m \in \calM_{\bp_i}^{\rm rx}} {\rm tr}\left(\bm{\Xi}_{i,m} \ddot{\bD}_{i,m}\bR_{i,m} \ddot{\bD}_{i,m}^{\rm H}\bm{\Xi}_{i,m}^{\rm H}\right)}{\ds \sum\nolimits_{m \in \calM_{\bp_i}^{\rm rx}}r_{i,m}},
    \end{aligned}
    \label{eq:gamma_detection}
\end{equation}
with $\bR_{i,m}= \bbE[\bm{\alpha}_{i,m}\bm{\alpha}_{i,m}^{\rm H}]$ the covariance matrix of $\bm{\alpha}_{i,m}$ (defined in Section~\ref{sec:7}) and $\bm{\Xi}_{i,m} = \bU_{i,m}^{\rm H} \bm{\Psi}_m^{-0.5}$ for convenience. Note that for performance evaluation, all terms will be computed w.r.t. the actual positions of the targets (rather than the radar cell position within the scanning grid).

\subsection{Tracking}
Similar to before, in the case of tracking, we are interested in sensing targets at (estimated) positions $\hat{\bp}_l$. However, instead of having widely separated range cells to avoid interference among echoes, the target locations are now not constrained. The reason is that, after scanning the whole area and sensing the targets in different time stamps during detection, we end up with sets of (possibly) overlapping targets to be tracked.

The received signal at AP $m$ is (cf. \eqref{eq:y_tilde_m})
\begin{equation}    
    \tilde{\by}_m[t] = \sum_{l =1}^L a\left(\hat{\bp}_l\right) \sum_{m' \in \calM^{\rm tx}} \bH_{l,m,m'} \bs_{m'}[t] + \tilde{\bg}_m[t] + \tilde{\bz}_m[t],
    \label{eq:y_tilde_m_tracking}
\end{equation}
where the echoes from other (close) targets cannot be disregarded. This means multi-hypothesis testing is needed, whose formulation is generally cumbersome and is tackled via sequential or iterative (non-scalable) algorithms. 

Instead, we advocate for the worst-case scenario (all non-desired echoes are received, i.e., $a(\hat{\bp}_{l'}) = 1$ $\forall l' \neq l$) to have a robust and low-complexity solution. Following the steps from the previous subsection, we have the binary test
\begin{equation}
    \left\{\begin{array}{ll}
    H_1: & \ddot{\by}_m = \ds \ddot{\bD}_{l,m} \bm{\alpha}_{l,m} + \ddot{\bh}_{l,m} + \ddot{\bg}_m + \ddot{\bz}_m \\
    H_0: & \ddot{\by}_m = \ds \ddot{\bh}_{l,m} + \ddot{\bg}_m + \ddot{\bz}_m,
    \end{array} \right.    
    \label{eq:HT_tracking}
\end{equation}
with $\ddot{\bh}_{l,m} = \sum_{l'\neq l} \ddot{\bD}_{l',m} \bm{\alpha}_{l',m}$ the interference coming from the other targets\footnote{In practice, echoes coming from far away targets can still be ignored.}. It is important to highlight that, although we are now focusing on the estimated position $\hat{\bp}_l$ for designing the beamformer/receiver processing, at the detection stage we assume this location to be perfectly known (constant), and later we model it according to \eqref{eq:Position_tracking} for the performance analysis.

Based on that, the GLRT reads as
\begin{equation}
    \frac{\ds {\rm max}_{\left\{\bm{\alpha}_{l,m}\right\}_{\forall l}} f\left(\ddot{\by}_m \vert H_1, \left\{\bm{\alpha}_{l,m}\right\}_{\forall l},\ddot{\bQ}_m \right)}{\ds {\rm max}_{\left\{\bm{\alpha}_{l',m}\right\}_{l' \neq l}} f\left(\ddot{\by}_m \vert H_0, \left\{\bm{\alpha}_{l',m}\right\}_{l' \neq l},\ddot{\bQ}_m \right)} \test \delta_l,
    \label{eq:GLRT_tracking}
\end{equation}
where the set of RCS vectors in the numerator includes the desired target response $\bm{\alpha}_{l,m}$ as well as the contribution of the interfering echoes $\left\{\bm{\alpha}_{l',m}\right\}_{l' \neq l}$ from the denominator.

Accordingly, \eqref{eq:GLRT_tracking} can be equivalently written as
\begin{equation}
    \frac{\ds {\rm exp}\left({\rm min}_{\left\{\bm{\alpha}_{l,m}\right\}_{\forall l}}  \sum\nolimits_{l = 1}^L\Theta\left(\bm{\alpha}_{l,m}\right)\right)}{\ds {\rm exp}\left({\rm min}_{\left\{\bm{\alpha}_{l',m}\right\}_{l' \neq l}}  \sum\nolimits_{l' \neq l}\Theta\left(\bm{\alpha}_{l',m}\right)\right)} \test \delta_l,
\end{equation}
where $\Theta(\bm{\alpha}_{l,m})$ is given in \eqref{eq:Theta} at the top of the next page.

\begin{figure*}[t]
    \begin{equation}                
        \Theta\left(\bm{\alpha}_{l,m}\right) = \bm{\alpha}_{l,m}^{\rm H}\ddot{\bD}_{l,m}^{\rm H}\bm{\Psi}_m^{-1}\ddot{\bD}_{l,m}\bm{\alpha}_{l,m} - 2\Re\left\{\bm{\alpha}_{l,m}^{\rm H}\ddot{\bD}_{l,m}^{\rm H}\bm{\Psi}_m^{-1} \ddot{\by}_m \right\}.
    \label{eq:Theta}
    \end{equation} 
    \hrule
\end{figure*}

Unlike before, now we have separate optimizations for the numerator and denominator. Fortunately, both problems are convex and admit a solution resembling the structure in \eqref{eq:alpha_opt_detection}. Also, since there are no cross-products and variables are decoupled ($\bm{\alpha}_{l,m}$ does not affect $\bm{\alpha}_{l',m}$), the GLRT becomes
\begin{equation}
        \frac{\ds {\rm exp}\left(\sum\nolimits_{l = 1}^L {\rm min}_{\bm{\alpha}_{l,m}}  \Theta\left(\bm{\alpha}_{l,m}\right)\right)}{\ds {\rm exp}\left(\sum\nolimits_{l' \neq l} {\rm min}_{\bm{\alpha}_{l',m}} \Theta\left(\bm{\alpha}_{l',m}\right)\right)} \test \delta_l,
\end{equation}
and whose individual solutions are
\begin{equation}
    \hat{\bm{\alpha}}_{l,m} = \left( \ddot{\bD}_{l,m}^{\rm H} \bm{\Psi}_m^{-1} \ddot{\bD}_{l,m}\right)^{-1} \ddot{\bD}_{l,m}^{\rm H} \bm{\Psi}_m^{-1} \ddot{\by}_m.
    \label{eq:alpha_opt_tracking}
\end{equation}

Finally, when substituting the optimal values, the CPU fuses the GLRTs from the APs $m \in \calM_{\hat{\bp}_l}^{\rm rx} = \calM_{\hat{\bp}_l} \cap \calM^{\rm rx}$ as
\begin{equation}
    \sum\nolimits_{m \in \calM_{\hat{\bp}_l}^{\rm rx}} \norm{\bU_{l,m}^{\rm H} \bm{\Psi}_m^{-0.5}\ddot{\by}_m}^2 \test \ln \left(\delta_l\right),
\end{equation}
where $\bU_{l,m} \in \mathbb{C}^{N \tau_s \times r_{l,m}}$ contains the left-singular vectors of the matrix $\bm{\Psi}_m^{-0.5}\ddot{\bD}_{l,m}$, with $r_{l,m} = {\rm rank}(\bm{\Psi}_m^{-0.5}\ddot{\bD}_{l,m}) \leq \vert \calM^{\rm tx} \vert$. Different from \eqref{eq:GLRT_detection_CPU}, the impact of the rest of the echoes is indeed included in the observations $\ddot{\by}_m$.

As a result, instead of an SCNR, we now have the following signal-to-interference-plus-clutter-and-noise-ratio (SICNR):
\begin{equation}
    \begin{aligned}
        \gamma_{\hat{\bp}_l}  &= \frac{\ds \bbE \left[\sum\nolimits_{m \in \calM_{\hat{\bp}_l}^{\rm rx}} \norm{ \bU_{l,m}^{\rm H} \bm{\Psi}_m^{-0.5}\ddot{\bD}_{l,m} \bm{\alpha}_{l,m}}^2 \right]}
        {\ds \bbE \left[\sum\nolimits_{m \in \calM_{\hat{\bp}_l}^{\rm rx}} \norm{ \bU_{l,m}^{\rm H} \bm{\Psi}_m^{-0.5} \left(\ddot{\bh}_{l,m} + \ddot{\bg}_m + \ddot{\bz}_m\right)}^2 \right]} \\
        &= \frac{\ds \sum\nolimits_{m \in \calM_{\hat{\bp}_l}^{\rm rx}} {\rm tr}\left(\bm{\Xi}_{l,m} \ddot{\bD}_{l,m}\bR_{l,m} \ddot{\bD}_{l,m}^{\rm H} \bm{\Xi}_{l,m}^{\rm H}\right)}{\ds \sum\nolimits_{m \in \calM_{\hat{\bp}_l}^{\rm rx}} r_{l,m} + \xi_{l,m}} , 
    \end{aligned}
    \label{eq:gamma_tracking}
\end{equation}
with $\bm{\Xi}_{l,m} = \bU_{l,m}^{\rm H} \bm{\Psi}_m^{-0.5}$ the receive combining (or tracking) matrix. Additionally, due to the independence among different RCSs $\bm{\alpha}_l$, we have
\begin{equation}
    \xi_{l,m} = \sum\nolimits_{l' \neq l} {\rm tr}\left(\bm{\Xi}_{l,m} \ddot{\bD}_{l',m}\bR_{l',m} \ddot{\bD}_{l',m}^{\rm H} \bm{\Xi}_{l,m}^{\rm H}\right).
\end{equation}

\section{Power Control Rules} \label{sec:6}
\subsection{Unscalable Policy: Minimum-Rate Maximization} \label{sec:6.1}
The GLRT receivers presented in Section~\ref{sec:5} are derived for a given power allocation. This means that the related detection metrics $\gamma_{\bp_i}$ ($\gamma_{\hat{\bp}_l}$), whose equivalency with the actual detection probability $p_{\rm d}$ will be studied later, are used for performance evaluation at the experimental stage once the set of coefficients $\eta_{k,m}$ and $\mu_{i,m}$ ($\mu_{l,m}$) are fixed \cite[Section~VIII]{behdad2024isac}. Note that, since the SCNR in \eqref{eq:gamma_detection} is a particularization of the SICNR in \eqref{eq:gamma_tracking} for $\xi_{l,m} = 0$, we concentrate on the tracking phase.

To design the coefficients, one can obtain communication-like metrics for the sensing. In a nutshell, according to \cite[Section~II]{lu2024isac}, we can establish the following (ergodic) mutual information over resource blocks:
\begin{equation}
    \bar{R}_l = \frac{\ds \tau_s}{\ds \tau_c} B \log_2\left(1 + \bar{\gamma}_l\right),
\end{equation}
with $\bar{\gamma}_l$ the “effective” signal-to-interference ratio (SIR) for object tracking. Note that clutter and noise are not included because the GLRT receive processing $\bm{\Xi}_{l,m}$ already whiteness these two processes (it acts as an MMSE filter). In other words, since interference is the ultimate limiting factor (like in most wireless networks), we focus on mitigating its impact.

As derived in the Appendix, information-theoretic reasonings applied to \eqref{eq:HT_tracking} yield the following SIR:
\begin{equation}
    \bar{\gamma}_l = \ds \sum\nolimits_{i= 1}^{\vert \calM^{\rm tx} \vert} \bb_i^{\rm T} \bar{\bF}_{l,i} \bb_i  \times \left(\ds \sum\nolimits_{i= 1}^{\vert \calM^{\rm tx} \vert} \bb_i^{\rm T} \tilde{\bF}_{l,i} \bb_i\right)^{-1},    
\label{eq:gamma_bar_l}
\end{equation} 
where the matrices $\bar{\bF}_{l,i}$ and $\tilde{\bF}_{l,i}$ are defined in the Appendix and the vector $\bb_i$ contains the (squared root) power coefficients of AP $i$ (cf. \eqref{eq:d_concatenated}). For ease of notation, in the sequel, one might write the SIR as $\bar{\gamma}_l = \bar{\calA}_l/\bar{\calB}_l$.

At this point, we can conceive a QoS-based power control to boost performance within the CF-mMIMO network. That is, by denoting $\bm{\eta} = [\eta_{1,1},\ldots,\eta_{K,M}]$ and $\bm{\mu} = [\mu_{1,1},\ldots,\mu_{L,M}]$ the vectors of power coefficients, the following optimization problem can be formed:
\begin{align} 
    \label{eq:Power_control} \underset{\bm{\eta},\bm{\mu}}{\rm max} \, \underset{k}{\rm min} \, & \log_2\left(1 + \gamma_k \right) \\
    {\rm s.t.} \quad & C1: \eta_{k,m}, \mu_{l,m} \geq 0 \quad \forall k,l,m \nonumber\\
    &C2:  \sum\nolimits_{k \in \calK_m} \eta_{k,m} + \sum\nolimits_{l \in \calL_m}\mu_{l,m} \leq P_m \quad \forall m  \nonumber \\    
    &C3:  \bar{\gamma}_l \geq \bar{\gamma}_0 \quad \forall l, \nonumber    
\end{align}
where $C1$ and $C2$ constrain the power budget for each AP, whereas $C3$ ensures all targets are detected with a sufficiently high SIR ($\bar{\gamma}_0$) to attain a good probability of detection (as long as the probability of false alarm is maintained \cite{richards2023principles}). Recall that, throughout the paper, we assumed unit norm beamformers (which is why their power is not included). Besides, given its complexity, the solution to the problem in \eqref{eq:Power_control} will only serve as a benchmark during the simulations (cf. \cite{interdonato2023coexistence}).

Like in most CF works, the objective in \eqref{eq:Power_control} is nonconvex (ratio of convex functions). However, taking a closer look, one can see it can also be written as a second-order cone (SOC) constraint \cite{demirhan2024cellfree}. More specifically, the problem above can be equivalently reformulated into standard epigraph form:
\begin{align}
    \label{eq:Power_control_SOC} \underset{\bm{\zeta},\bm{\nu},\gamma}{\rm max} \quad & \gamma \\
    {\rm s.t.} \quad & C1: \zeta_{k,m}, \nu_{l,m} \geq 0 \quad \forall k,l,m \nonumber\\
    &C2:  \sum\nolimits_{k \in \calK_m} \zeta_{k,m}^2 + \sum\nolimits_{l \in \calL_m}\nu_{l,m}^2 \leq P_m \quad \forall m \nonumber \\    
    &C3:  \bar{\gamma}_l \geq \bar{\gamma}_0 \quad \forall l  \nonumber\\        
    &C4: \norm{\bm{\varrho}_k} \leq \sqrt{1 + \frac{1}{\gamma}} \sum_{m \in \calM_k} \zeta_{k,m} \sqrt{{\rm tr}\left(\bm{\Phi}_{k,m}\right)} \quad \forall k, \nonumber    
\end{align}
where $\bm{\zeta} = [\zeta_{1,1},\ldots,\zeta_{K,M}]$ and $\bm{\nu} = [\nu_{1,1},\ldots,\nu_{L,M}]$, with $\zeta_{k,m} = \sqrt{\eta_{k,m}}$ and $\nu_{l,m} = \sqrt{\mu_{l,m}}$ the new design variables. For the SOC $C4$, we have defined the auxiliary variable $\bm{\varrho}_k = [\varpi_{k,1},\ldots,\varpi_{k,K},\bm{\varepsilon}_k,\bm{\vartheta}_k,\sigma_z]$ with
\begin{equation}
    \varpi_{k,j} = \bm{\pi}_j^{\rm H} \bm{\pi}_k  \sum\nolimits_{m \in \calM_j} \zeta_{j,m} \frac{\ds {\rm tr}\left(\bC_{k,m} \bm{\Lambda}_{j,m}\right)}{\ds \sqrt{{\rm tr}(\bm{\Phi}_{j,m})}},
\end{equation}
while the vectors $\bm{\varepsilon}_k$ and $\bm{\vartheta}_k$ contain the elements
\begin{equation*}
    \begin{aligned}
        \varepsilon_{k,j,m'} &= \zeta_{j,m'} \sqrt{\frac{\ds {\rm tr}\left(\bC_{k,m'} \bm{\Phi}_{j,m'}\right)}{\ds {\rm tr}(\bm{\Phi}_{j,m'})}}, \\
    \vartheta_{k,l,m} &= \nu_{l,m} \sqrt{{\rm tr}\left(\bC_{k,m}\bW_{l,m}\right)},    
    \end{aligned}    
\end{equation*}
from within the sets $m' \in \calM_j$, $m \in \calM^{\rm tx}$, and $l \in \calL_m$.

As shown in the Appendix, the SIR $\bar{\gamma}_l$ is also a ratio of convex functions w.r.t. the new coefficients $\zeta_{k,m}$ and $\nu_{i,m}$. Unfortunately, their coupling within $\bar{\gamma}_l$ leads to a nonconvex constraint $C3$ that is usually tackled via SCAs (successive convex approximations) \cite{liesegang2024emf}. This allows us to find a local optimum under mild assumptions.

In summary, for a given $\gamma$, the optimization is decomposed into a sequence of subproblems to be solved iteratively. Each of them must have a global optimum for guaranteeing convergence, which means the numerator in $\bar{\gamma}_l$ must be approximated by a surrogate function \cite{sun2017mm}. On top of that, a bisection search can be applied to derive the optimal $\gamma$ \cite[Algorithm~2]{ngo2017cell}.

Among others, a common strategy is to linearize the function $\bar{\calA}_l$ so that the constraint $C3$ is convexified. In particular, when applying the first-order Taylor series expansions at the previous feasible point, we can obtain the lower bound:
\begin{equation}
    \begin{aligned}
        \bar{\calA}_l\left(\bb\right) &\geq \bar{\calA}_l\left(\bb^{(u - 1)}\right) + \left(\nabla \bar{\calA}_l\left(\bb^{(u - 1)}\right)\right)^{\rm T}\left(\bb - \bb^{(u - 1)}\right) \\
        &\triangleq \tilde{\calA}_l\left(\bb,\bb^{(u - 1)}\right),    
    \end{aligned}    
    \label{eq:Taylor}
\end{equation}
with $\bb = [\bm{\zeta}^{\rm T} \bm{\nu}^{\rm T}]^{\rm T}$ the concatenation of all the “amplitude” coefficients, $u$ the iteration index, and the gradient
\begin{equation}
    \nabla \bar{\calA}_l\left(\bb\right) = 2 \left(\sum\nolimits_{i= 1}^{\vert \calM^{\rm tx} \vert} \bP_i^{\rm T}\Re \lbrace \bar{\bF}_{l,i} \rbrace \bP_i\right) \bb,
\end{equation}
where $\bP_i \in \{0,1\}^{(K + L) \times M(K + L)}$ is introduced for notational consistency, i.e., $\bP_i \bb \equiv \bb_i$ (the coefficients from AP $i$).

Accordingly, at the $u$-th iteration, we will need to solve the following feasibility problem \cite{ngo2017cell}:
\begin{equation}
\begin{aligned}
        \textrm{find} \quad & \bb \\
        \textrm{s.t.} \quad & C1,C2,C4 \\
        &C3: \bar{\gamma}_0 \bar{\calB}_l \left(\bb \right) + \tilde{\calA}_l\left(\bb,\bb^{(u - 1)}\right) \leq 0 \quad \forall l,
    \end{aligned}
    \label{eq:Feasibility_ISAC}      
\end{equation}
where we replaced the original difference of convex functions with the single convex function from \eqref{eq:Taylor}. 

This way, for each $\gamma$ fixed by bisection, we end up with a series of subproblems that are worst-case scenarios (more restrictive constraints) though globally solvable with standard numerical routines like CVX \cite{cvx2020}. The procedure is iterated until convergence to a local optimum, as summarized in Algorithm~\ref{alg:SCA}, with $\mathcal{S} = \{\bb \, \vert \, C1,\ldots,C4\}$ the feasible set.

\subsection{Scalable Policy: Generalized Fractional Allocation} \label{sec:4.2}
The previous method faces expensive computations, making the resulting allocation unfeasible for large networks. Following the fractional power control (FPC) idea \cite{demir2021foundations}, we can design fully scalable schemes by generalizing them to ISAC.

With the above considerations, the FPC coefficients are
\begin{equation}
    \eta_{k,m} = p_m\rho_{k,m}^{\kappa_{\rm c}}, \quad \mu_{l,m} = p_m\lambda_{l,m}^{\kappa_{\rm s}},
    \label{eq:scalable_distributed}
\end{equation}
where $\rho_{k,m}$ is the LSF coefficient between (transmit) AP $m$ and UE $k$, $\lambda_{l,m}$ is the LSF from AP $m$ to target in position $\hat{\bp}_l$, and $\kappa_{\rm c}$ ($\kappa_{\rm s}$) are parameters to tune the distribution of power: 
\begin{itemize}
    \item $\kappa = 1$ prioritizes good channel qualities, also known as proportional power control (PPC).
    \item $\kappa = 0$ provides equal service in the network, i.e., uniform power control (UPC).
    \item $\kappa = -1$ resembles a fairer policy that helps links with poor propagation.
\end{itemize}

Hence, network operators can choose the combination of the tuple $(\kappa_{\rm c},\kappa_{\rm s})$ that better suits their needs. Finally, $p_m$ scales the resulting coefficients to satisfy the power constraints, i.e.,
\begin{equation}
    p_m = \frac{\ds P_m}{\ds \sum\nolimits_{k \in \calK_m} \rho_{k,m}^{\kappa_c} + \sum\nolimits_{l \in \calL_m}\lambda_{l,m}^{\kappa_{\rm s}}}.
    \label{eq:p_m}
\end{equation}

However, when using different values for $\kappa_{\rm c}$ and $\kappa_{\rm s}$, the gap between the orders of magnitude of the coefficients might become very large, e.g., for $\kappa_{\rm c} = 0$ (UPC) and $\kappa_{\rm s} = 1$ (PPC), we will have $\rho_{k,m}^{\kappa_{\rm c}} = 1$ and $\lambda_{l,m}^{\kappa_{\rm s}} \ll 1$ (provided that normally the LSF coefficients are small, i.e., $\rho_{k,m},\lambda_{l,m} < 1$). 

To overcome this issue, we must introduce some sort of normalization. A feasible option is to consider
\begin{equation}
    \eta_{k,m} = p_m c_m \rho_{k,m}^{\kappa_{\rm c}}, \quad \mu_{l,m} = p_m s_m \lambda_{l,m}^{\kappa_{\rm s}},
\end{equation}
with the newly introduced (constant) factors
\begin{equation}
    c_m = \frac{1}{\ds {\rm max}_k \, \rho_{k,m}^{\kappa_{\rm c}}}, \quad s_m = \frac{1}{\ds {\rm max}_l \, \lambda_{l,m}^{\kappa_{\rm s}}}.
\end{equation}

Accordingly, \eqref{eq:p_m} is rewritten as
\begin{equation}
    p_m = \frac{\ds P_m}{\ds c_m \sum\nolimits_{k \in \calK_m} \rho_{k,m}^{\kappa_c} + s_m \sum\nolimits_{l \in \calL_m}\lambda_{l,m}^{\kappa_{\rm s}}}.
\end{equation}

\section{Numerical Simulations} \label{sec:7}
Throughout all experiments, we consider a deployment area of $0.5$ km\textsuperscript{2}, where $L = 4$ targets are uniformly deployed with heights ranging between $20$ m and $100$ m. During the detection phase, this area is equally divided into $S = 9$ sensing regions. UEs and APs are also randomly located within the whole area but at fixed heights of $1.65$ m and $10$ m, respectively. We assume $K = 16$ UEs linked to $4$ APs from a total of $M = 16$ equipped with uniform linear arrays of $N = 4$ antennas. This association is based on the LSF coefficients, i.e., each UE is served by the $N$ APs with the highest values.

We employ the micro-urban scenario from \cite{3GPP36814}, with $P_m = 2$ W, $\bar{\eta}_k = 0.1$ mW, $\sigma_z^2 = N_o B$, $N_o = -174$ dBm/Hz, and $B = 20$ MHz. The carrier frequency is $2$ GHz and the coherence block contains $\tau_c = 50$ symbols (which is equivalent to typical 5G coherence bandwidths of $100$ kHz and time slots of $0.5$ ms). The target response is modeled as ${\alpha}_{l,m,m'} \sim \calCN(0,\sigma_{\alpha}^2)$, where, unless otherwise specified, $\sigma_{\alpha}^2 = 10$ dBsm (decibel relative to one square meter) and a Gaussian-shaped correlation based on the APs angle of view. The tracking position error in \eqref{eq:Position_tracking} is given by $\bm{\varepsilon}_l \sim \calN(0,\sigma_{\varepsilon}^2\bI_2)$. Besides, for each region (or target), we consider $M_{\rm tx} = 4$ transmit APs and $M_{\rm rx} = 1$ receive AP. These APs are chosen as the closest to the inspected radar cells or potential targets. Finally, the number of sensing samples is generally $\tau_s = \tau_c = 50$ samples.

\begin{algorithm}[t]
\begin{algorithmic}[1]    
    \State Choose convergence tolerance thresholds $\epsilon_1, \epsilon_2 > 0$    
    \State Select feasible bisection bounds $\gamma_{\rm min}$ and $\gamma_{\rm max}$
    \State Initialize coefficients $\bb^{(0)} \in \mathcal{S}$  
    \While{$\gamma_{\rm max} - \gamma_{\rm min} > \epsilon_1$}
        \State Set $\gamma = (\gamma_{\rm max} + \gamma_{\rm min})/2$ and $u = 1$
        \While{$\| \bb - \bb^{(u - 1)} \|^2/ \| \bb \|^2 > \epsilon_2$}
            \State Solve \eqref{eq:Feasibility_ISAC} with $\bb^{(u - 1)}$ to find $\bb$ using \cite{cvx2020}
    		\State Set $u = u + 1$ and $\bb^{(u - 1)} = \bb$   
        \EndWhile
        \If{problem \eqref{eq:Feasibility_ISAC} is feasible}
            \State Set $\gamma_{\rm min} = \gamma$ and $\bb^{(0)} = \bb$        
        \Else
            \State Set $\gamma_{\rm max} = \gamma$
        \EndIf
    \EndWhile
\end{algorithmic}
\caption{SCA-based Solution with SOC}
\label{alg:SCA}
\end{algorithm}

Following the discussion in \cite{behdad2024isac}, we further assume some clutter parts are “virtually” canceled thanks to other sources of channel information, e.g., the presence of permanent obstacles. That is, the power of the AP-AP channels $\bar{G}_{m,m'}$ will be scaled by a factor $\varsigma \in [0,1]$ depending on the degree of additional knowledge. As observed later, this new control parameter will open the door to deeper insights. Despite that, we initially fix $\varsigma = 10^{-2}$ for the main baseline configuration.

The system performance is studied via the CDF (cumulative distribution function) of the UE data rate and the sensing SCNR (or SICNR), calculated by averaging over 100 random UE and target locations (each with 1000 fading realizations). 

\subsection{Performance Evaluation: Detection}
Prior to analyzing the UE data rates and sensing SCNR, we start this subsection by presenting the probability of miss $p_{\rm M} = 1 - p_{\rm D}$, with $p_{\rm D}$ the probability of detection. This will justify the adoption of SCNR as a valuable figure of merit.

\begin{figure}[t]    
    \centerline{\includegraphics[scale = 1]{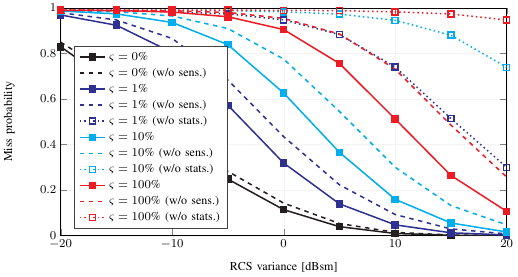}}    
    \caption{Miss probability $p_{\rm M}$ vs. RCS variance $\sigma_{\alpha}^2$ for different clutter percentages $\varsigma$ with (full) ISAC, w/o sensing, and w/o clutter stats.}    
    \label{fig:2}    
    \vspace{2mm}
    \begin{subfigure}[b]{0.24\textwidth}        
        \centerline{\includegraphics[scale = 1]{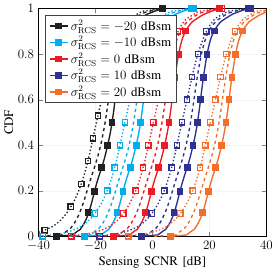}}
    \end{subfigure}
    \hfill
    \nextfloat
    \begin{subfigure}[b]{0.24\textwidth}        
        \centerline{\includegraphics[scale = 1]{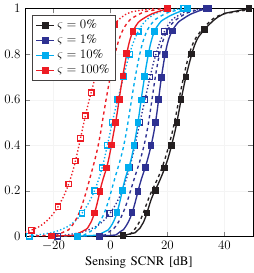}}
    \end{subfigure}    
    \caption{CDF of the SCNR vs. $\sigma_{\alpha}^2$ (left) and $\varsigma$ (right). Dashed (dotted) lines indicate the absence of sensing-dedicated signals (clutter stats.).}
    \label{fig:3}        
\end{figure}

Accordingly, in Fig.~\ref{fig:2}, we present the miss probability $p_{\rm M}$ w.r.t. the RCS variance $\sigma_{\alpha}^2$ and different values of $\varsigma$ (in percentage). This is done for both the cases of no sensing-dedicated signals, i.e., $\mu_{l,m} = 0$, and ISAC (with UPC). At a glance, one can state that ISAC plays a crucial role in the detection (success degrades when only data signals are used, especially for large $\varsigma$). Besides that, the performance behaves as expected: the number of misses decreases with $\sigma_{\alpha}^2$ and increases with $\varsigma$. In that sense, we also include the results of a GLRT based solely on the noise statistics. This detector is shown to exhibit a major loss, stressing the importance of designing clutter-aware detection algorithms.

The same pattern is also observed in Fig.~\ref{fig:3}, where the CDF of the SCNR $\gamma_{\bp_i}$ is illustrated when varying $\sigma_{\alpha}^2$ and $\varsigma$. This surely verifies that we can safely replace the miss probability with the ratio $\gamma_{\bp_i}$. Notably, for the reference values $\sigma_{\alpha}^2 = 10$ dBsm and $\varsigma = 10^{-2}$, results indicate that $\gamma_{\bp_i} \geq 0$ dB is sufficient for good target detection (e.g., $p_{\rm M} \leq 0.1$).

\begin{figure}[t]    
    \vspace{-0.1mm}
    \begin{subfigure}[b]{0.24\textwidth}            
        \centerline{\includegraphics[scale = 1]{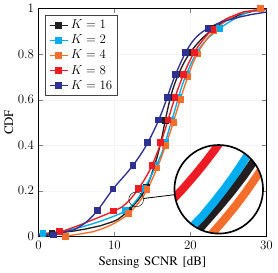}}        
        \vspace{1mm}
    \end{subfigure}
    \hfill    
    \begin{subfigure}[b]{0.24\textwidth}                
        \centerline{\includegraphics[scale = 1]{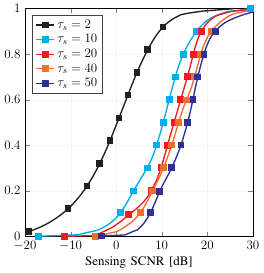}}
        \vspace{1mm}
    \end{subfigure}    
    \caption{CDF of the SCNR vs. number of UEs $K$ (left) and number of sensing samples $\tau_s$ (right). The solid blue line indicates the baseline.}
    \label{fig:4}
    \vspace{2mm}
    \begin{subfigure}[b]{0.24\textwidth}        
        \centerline{\includegraphics[scale = 1]{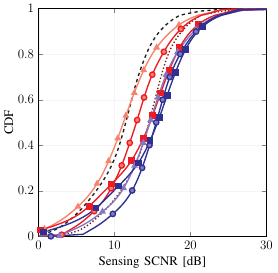}}
    \end{subfigure}
    \hfill
    \begin{subfigure}[b]{0.24\textwidth}        
        \centerline{\includegraphics[scale = 1]{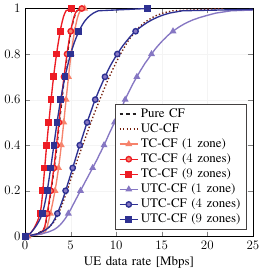}}
    \end{subfigure}    
    \caption{CDF of the SCNR (left) and data rate (right) under: pure CF (dashed), UC-CF (dotted), TC-CF, UTC-CF (solid) implementations.}
    \label{fig:5}       
\end{figure}

On the other hand, the SCNR versus (vs.) the number of UEs $K$ and number of sensing samples $\tau_s$ is pictured in Fig.~\ref{fig:4}. In brief, there is a clear trade-off w.r.t. $K$: too many UEs limit the sensing resources, whereas too few prevent their echoes' recycling. It seems $K = 4$ UEs provides the best compromise. Contrarily, higher $\tau_s$ always boosts performance, yet the gap is more significant when less than $10$ symbols are used.

Fig.~\ref{fig:5} provides evidence of the merits of the proposed user-centric and target-centric approach. Here, indeed, we show the evaluation of the SCNR and the data rate for different CF implementations: the user-and-target-centric (UTC); the user-centric (UC), where all targets are sequentially detected by all APs in a non-scalable way; the target-centric (TC), where all UEs are served by all APs in a non-scalable way; and pure CF, where both sensing and communication tasks are non-scalable. In all target-centric cases, we also modify the number of sensing regions $S$. 
Overall, the proposed UTC architecture outperforms the rest of the schemes, which motivates the need for scalable approaches. The choice of $S$ is, however, unclear. For data transmission, the lower, the better (more resources are available for the UEs). For sensing, the previous trade-off strikes again: $S = 9$ implies unnecessary beams that waste energy, whereas $S=1$ leads to insufficient received power for detection. A good balance is found at $S = 4$. Still, in light of its lower sensitivity (compared to that of the rate), $S = 1$ could be enough for sensing. 

Finally, the outcomes of the generalized FPC are depicted in Fig.~\ref{fig:6} w.r.t. relevant $(\kappa_c,\kappa_s)$ tuples (the minimax policy is evaluated in the upcoming subsection to avoid redundancy). We then concentrate on opportunistic values $\kappa \in [0,1]$, which are usually preferred in the DL \cite[(7.47)]{demir2021foundations}. Notably, one can see that modifying $\kappa$ is always beneficial, especially in terms of data rate. For any $\kappa_c$, using larger $\kappa_s$ yields considerable gains (sensing beams become more directive). As a result, given its slight impact on SCNR, the combination $\kappa_c = 0$ and $\kappa_s = 1$ sounds appropriate to ensure good QoS for all UEs.

\subsection{Performance Evaluation: Tracking}
We now investigate the system performance when tracking is performed: as a result, interference from nearby other targets may be experienced during sensing; moreover, the effects of the position error\footnote{Note that, in the detection, we have also introduced the mismatch between the sensed and target positions ($\bp_i$ is the center of the inspected range cell).} $\bm{\varepsilon}_l$ with $\sigma_{\varepsilon} = 10^{-2}$ are also considered. 

\begin{figure}[t]    
    \begin{subfigure}[b]{0.24\textwidth}        
        \centerline{\includegraphics[scale = 1]{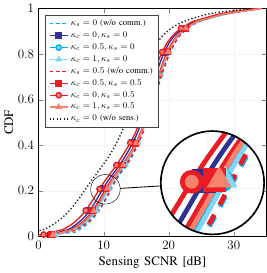}}
    \end{subfigure}
    \hfill
    \begin{subfigure}[b]{0.24\textwidth}    
        \centerline{\includegraphics[scale = 1]{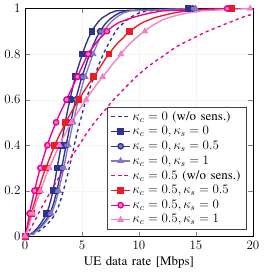}}
    \end{subfigure}    
    \caption{CDF of the SCNR (left) and data rate (right) vs. FPC. Dashed and dotted lines indicate the absence of sensing or communication.}
    \label{fig:6}
    \vspace{2mm}    
    \begin{subfigure}[b]{0.24\textwidth}        
        \centerline{\includegraphics[scale = 1]{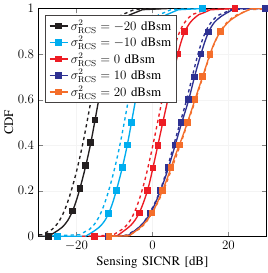}}
    \end{subfigure}
    \hfill
    \begin{subfigure}[b]{0.24\textwidth}        
        \centerline{\includegraphics[scale = 1]{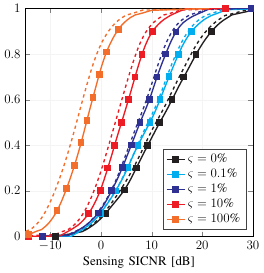}}
    \end{subfigure}    
    \caption{CDF of the SICNR vs. RCS variance $\sigma_{\alpha}^2$ (left) and clutter percentage $\varsigma$ (right). Dashed lines indicate the absence of sensing.}
    \label{fig:7}   
\end{figure}

We begin by illustrating, in Fig.~\ref{fig:7}, the CDF of the SICNR w.r.t. the RCS variance $\sigma_{\alpha}^2$ and clutter power factor $\varsigma$. Remarkably, the advantage of raising $\sigma_{\alpha}^2$ decreases with higher values. This is reasonable since it also augments the power of the other echoes in the denominator from $\gamma_{\hat{\bp}_l}$. In addition, the larger the RCS variance is, the smaller the loss of no sensing is (the system becomes more interference-limited). Likewise, the deterioration with increasing $\varsigma$ diminishes as the performance is no longer governed by aggregate clutter plus noise. 

\begin{figure}[t]    
    \begin{subfigure}[b]{0.24\textwidth}        
        \centerline{\includegraphics[scale = 1]{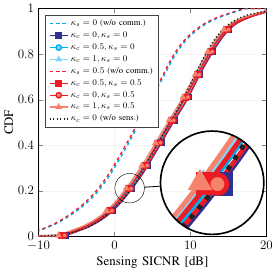}}
    \end{subfigure}
    \hfill
    \begin{subfigure}[b]{0.24\textwidth}        
        \centerline{\includegraphics[scale = 1]{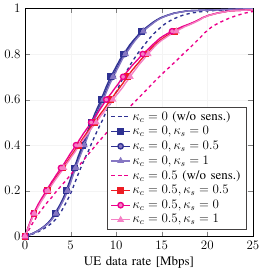}}
    \end{subfigure}    
    \caption{CDF of the SICNR (left) and data rate (right) vs. FPC. Dashed and dotted lines indicate the absence of sensing or communication.}
    \label{fig:8}
    \vspace{2mm}    
    \begin{subfigure}[b]{0.24\textwidth}        
        \centerline{\includegraphics[scale = 1]{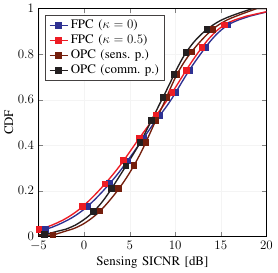}}
        \vspace{0.25mm}
    \end{subfigure}
    \hfill
    \begin{subfigure}[b]{0.24\textwidth}        
        \centerline{\includegraphics[scale = 1]{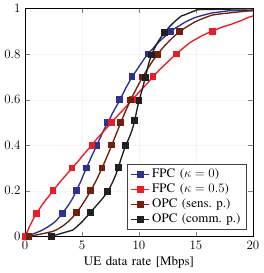}}
        \vspace{0.25mm}
    \end{subfigure}    
    \caption{CDF of the SICNR (left) and data rate (right) vs. FPC and OPC. In the latter, the abbreviation \textit{p.} stands for \textit{prioritized} (design).}
    \label{fig:9}   
\end{figure}

Then, in Fig.~\ref{fig:8}, the SICNR and rate w.r.t. the FPC policies are shown. Similar to before, increasing $\kappa_s$ leads to higher data rates. Interestingly, though, removing communication implies stronger SICNR degradation than excluding sensing. Again, opposite to detection (cf. Fig.~\ref{fig:6}), this is because \textit{sensing} interference now dominates over clutter and noise. More precisely, since communication beams point towards ground UEs (not aerial targets), their contribution to interference is less severe. This suggests tracking is possible with only data signals, even though the situation could change with interference-mitigation techniques (e.g., joint decoding of all targets) and with more advanced beamforming structures. Due to the lack of space, the extension with such schemes is left for future studies.

Last, the (unscalable) minimax solution, labeled as optimal power control (OPC), and the (scalable) FPC policies with fair and opportunistic $(\kappa_c,\kappa_s) \in \{0,0.5\}$, are compared in Fig.~\ref{fig:9}. Recall that, given its prohibitive complexity, the OPC can only serve as a benchmark. Besides, we further distinguish between: (i) \textit{communication-prioritized} OPC, where the SIR threshold $\bar{\gamma}_0$ in \eqref{eq:Power_control} is kept constant and equal to some predefined value; and (ii) \textit{sensing-prioritized} OPC, where we constantly increase $\bar{\gamma}_0$ until the problem becomes unfeasible. This results in more resources dedicated to UEs or targets, respectively. Unsurprisingly, both designs surpass FPC performance (although slightly in terms of SICNR). In addition, experiments reveal the cost-benefit in communication-prioritized OPC overtakes that in sensing-prioritized OPC. In other words, the improvement in data rate is larger than the loss in SICNR. This indeed confirms the premise that communication should eventually prevail in the trade-off between targets and UEs.

\section{Conclusions} \label{sec:8}
This paper has investigated a CF-mMIMO system where distributed APs performed ISAC tasks, detecting/tracking targets and communicating with UEs. A unified framework and signal model were developed for target detection and tracking in arbitrary positions. Using the GLRT technique, novel detection and tracking algorithms have been proposed to handle unknown target responses and interference. Moreover, scalable AP-UE and AP-target association rules have been introduced, considering multi-zone sensing. A scalable power allocation mechanism has then been introduced to extend the fractional power allocation to the ISAC paradigm. For benchmarking, a non-scalable power control optimization has been formulated and solved via SCA. Numerical results have validated the proposed design, highlighting the impact of target interference and trade-offs between communication and sensing realms.

\section*{Appendix: Effective Sensing SIR}
The effective SIR \(\bar{\gamma}_l\) for the \(H_1\)-signal in \eqref{eq:HT_tracking} extends \cite[(68)]{behdad2024isac} to multiple targets within the proposed decentralized architecture. The resulting SIR expression in \eqref{eq:effectiveSIR} is entirely new, with distinct derivations provided below.

Due to the independence among RCS, the SIR $\bar{\gamma}_l$ can be initially reported as per \eqref{eq:effectiveSIR} at the top of the next page. Recall that, during sensing, the transmit (probe) signals are known at the receiver side. In addition, for consistency between detection and tracking, we also assume the inspected position is perfectly known or estimated.

\begin{figure*}[t]    
    \begin{equation}
        \bar{\gamma}_l = \frac{\ds \bbE \left[\sum\nolimits_{m \in \calM_{\hat{\bp}_l}^{\rm rx}} \norm{\ddot{\bD}_{l,m} \bm{\alpha}_{l,m}}^2 \right]}{\ds
        \bbE \left[ \sum\nolimits_{m \in \calM_{\hat{\bp}_l}^{\rm rx}} \norm{\ddot{\bh}_{l,m}}^2\right]}    
         = \frac{\ds \sum\nolimits_{m \in \calM_{\hat{\bp}_l}^{\rm rx}} \sum\nolimits_{i= 1}^{\vert \calM^{\rm tx} \vert} \sum\nolimits_{j = 1}^{\vert \calM^{\rm tx} \vert} \bb_i^{\rm T} \bF_{l,m,i,j} \bb_j}{\ds \sum\nolimits_{m \in \calM_{\hat{\bp}_l}^{\rm rx}} \sum\nolimits_{l' \neq l} \sum\nolimits_{i= 1}^{\vert \calM^{\rm tx} \vert} \sum\nolimits_{j = 1}^{\vert \calM^{\rm tx} \vert} \bb_i^{\rm T} \bF_{l',m,i,j} \bb_j}   \label{eq:effectiveSIR}                
    \end{equation}            
    \hrule
\end{figure*}

Let us first rewrite the vectors $\bd_{l,m,m'}[t]$ in \eqref{eq:d_columns} as follows:
\begin{equation}
    \begin{aligned}
         \bd_{l,m,m'}[t]&= \sqrt{\beta_{l,m,m'}} \bA_{l,m,m'} (\bm{\Omega}_{m'}{\rm diag}\left(\bm{\upsilon}_{m'}[t]\right)\bm{\zeta}_{m'} \\
         &\quad + \bm{\Omega}_{0,m'}{\rm diag}\left(\bm{\chi}_{m'}[t]\right)\bm{\nu}_{m'}) \\
         &= \bm{\Upsilon}_{l,m,m'}[t]\bm{\zeta}_{m'} + \bm{\Upsilon}_{0,l,m,m'}[t]\bm{\nu}_{m'},
    \end{aligned}
\end{equation}
where $\bm{\Omega}_{m'}$ and $\bm{\Omega}_{0,m'}$ are the matrices of beamformers for communication and sensing at AP $m'$, i.e., their columns are the vectors $\bw_{k,m'}$ and $\bw_{0,m'}$ for $k \in \calK_m$ and $l \in \calL_m$, respectively. Accordingly, $\bm{\upsilon}_{m'}[t]$ and $\bm{\chi}_{m'}[t]$ contain the corresponding transmit symbols (cf. \eqref{eq:s_m} and \eqref{eq:s_0m}) and $\bm{\zeta}_{m'}$ and $\bm{\nu}_{m'}$ comprise the set of (squared-root) power coefficients.

By concatenation, the above becomes
\begin{equation}    
    \bd_{l,m,m'}[t] = \underbrace{\vphantom{\begin{bmatrix}\bm{\zeta}_{m'} \\ \bm{\nu}_{m'}\end{bmatrix}} \begin{bmatrix} \bm{\Upsilon}_{l,m,m'}[t] & \bm{\Upsilon}_{0,l,m,m'}[t] \end{bmatrix}}_{\triangleq \bS_{l,m,m'}[t]} \underbrace{\begin{bmatrix}\bm{\zeta}_{m'} \\ \bm{\nu}_{m'}\end{bmatrix}}_{\triangleq \bb_{m'}},
    \label{eq:d_concatenated}
\end{equation}
which allows us to express the numerator in \eqref{eq:effectiveSIR} as
\begin{equation}    
    \bar{\calA}_l = \sum\nolimits_{m \in \calM_{\hat{\bp}_l}^{\rm rx}} \sum\nolimits_{i= 1}^{\vert \calM^{\rm tx} \vert} \sum\nolimits_{j = 1}^{\vert \calM^{\rm tx} \vert} \bb_i^{\rm T} \bF_{l,m,i,j} \bb_j,      
\end{equation}
with $\bF_{l,m,i,j} = [\bR_{l,m}]_{i,j} \sum_{t = 1}^{\tau_s}\bS_{l,m,i}^{\rm H}[t] \bS_{l,m,j}[t]$, and the first term of the denominator in \eqref{eq:effectiveSIR} as 
\begin{equation}
    \bar{\calB}_l = \sum\nolimits_{m \in \calM_{\hat{\bp}_l}^{\rm rx}} \sum\nolimits_{l' \neq l} \sum\nolimits_{i= 1}^{\vert \calM^{\rm tx} \vert} \sum\nolimits_{j = 1}^{\vert \calM^{\rm tx} \vert} \bb_i^{\rm T} \bF_{l',m,i,j} \bb_j.
\end{equation}

Finally, it is easy to demonstrate that the SIR in \eqref{eq:effectiveSIR} is the ratio of two convex functions for uncorrelated RCS\footnote{Recall that this tractable metric is used only at the design stage, which is why it is reasonable to apply such simplification.}, i.e., $[\bR_{l,m}]_{i,j} = 0$ for $j \neq i$. In a few words, since the vector $\bb_i$ of coefficients is real-valued and the matrices $\bF_{l,m,i,i}$ are positive semi-definite by definition, it follows that $\bar{\calA}_l$ and $\bar{\calB}_l$ are convex w.r.t. $\bb_i$. As a result, under those assumptions, the final SIR is given by \eqref{eq:gamma_bar_l}, where $\bar{\bF}_{l,i} = \sum\nolimits_{m \in \calM_{\hat{\bp}_l}^{\rm rx}} \bF_{l,m,i,i}$ and $\tilde{\bF}_{l,i} = \sum\nolimits_{m \in \calM_{\hat{\bp}_l}^{\rm rx}} \sum\nolimits_{l' \neq l} \bF_{l',m,i,i}$. This completes the proof.

\bibliographystyle{IEEEtran}
\bibliography{references}

\begin{thebibliography}{10}
\providecommand{\url}[1]{#1}
\csname url@samestyle\endcsname
\providecommand{\newblock}{\relax}
\providecommand{\bibinfo}[2]{#2}
\providecommand{\BIBentrySTDinterwordspacing}{\spaceskip=0pt\relax}
\providecommand{\BIBentryALTinterwordstretchfactor}{4}
\providecommand{\BIBentryALTinterwordspacing}{\spaceskip=\fontdimen2\font plus
\BIBentryALTinterwordstretchfactor\fontdimen3\font minus \fontdimen4\font\relax}
\providecommand{\BIBforeignlanguage}[2]{{%
\expandafter\ifx\csname l@#1\endcsname\relax
\typeout{** WARNING: IEEEtran.bst: No hyphenation pattern has been}%
\typeout{** loaded for the language `#1'. Using the pattern for}%
\typeout{** the default language instead.}%
\else
\language=\csname l@#1\endcsname
\fi
#2}}
\providecommand{\BIBdecl}{\relax}
\BIBdecl

\bibitem{buzzi2024scalable}
S.~Buzzi, C.~D’Andrea, and S.~Liesegang, ``Scalability and implementation aspects of cell-free massive {MIMO} for {ISAC},'' in \emph{2024 IEEE 19th International Symposium on Wireless Communication Systems (ISWCS)}, 2024, pp. 1--6.

\bibitem{liu2022integrated}
F.~Liu \emph{et~al.}, ``Integrated sensing and communications: Toward dual-functional wireless networks for {6G} and beyond,'' \emph{IEEE J. Sel. Areas Commun.}, vol.~40, no.~6, pp. 1728--1767, 2022.

\bibitem{ngo2015cell}
H.~Q. Ngo \emph{et~al.}, ``Cell-free massive {MIMO}: Uniformly great service for everyone,'' in \emph{2015 IEEE 16th International Workshop on Signal Processing Advances in Wireless Communications (SPAWC)}, 2015, pp. 201--205.

\bibitem{demir2021foundations}
{\"O}.~T. Demir, E.~Bj{\"o}rnson, and L.~Sanguinetti, ``Foundations of user-centric cell-free massive {MIMO},'' \emph{Foundations and Trends{\textregistered} in Signal Processing}, vol.~14, no. 3-4, pp. 162--472, 2021.

\bibitem{buzzi2017cell}
S.~Buzzi and C.~D’Andrea, ``Cell-free massive {MIMO: User-centric} approach,'' \emph{IEEE Wireless Commun. Lett.}, vol.~6, no.~6, pp. 706--709, 2017.

\bibitem{buzzi2017user}
S.~Buzzi and C.~D'Andrea, ``User-centric communications versus cell-free massive {MIMO for 5G} cellular networks,'' in \emph{2017 21th International ITG Workshop on Smart Antennas (WSA)}, 2017, pp. 1--6.

\bibitem{buzzi2019using}
S.~Buzzi, C.~D’Andrea, and M.~Lops, ``Using massive {MIMO} arrays for joint communication and sensing,'' in \emph{2019 53rd Asilomar Conference on Signals, Systems, and Computers}, 2019, pp. 5--9.

\bibitem{liao2024powerallocation}
B.~Liao \emph{et~al.}, ``Power allocation for massive {MIMO-ISAC} systems,'' \emph{IEEE Trans. Wireless Commun.}, pp. 1--1, 2024.

\bibitem{lu2024isac}
S.~Lu \emph{et~al.}, ``Integrated sensing and communications: Recent advances and ten open challenges,'' \emph{IEEE Internet Things J.}, vol.~11, no.~11, pp. 19\,094--19\,120, 2024.

\bibitem{meng2024cooperative}
K.~Meng \emph{et~al.}, ``Cooperative {ISAC} networks: {Opportunities} and challenges,'' \emph{IEEE Wireless Commun.}, pp. 1--8, 2024.

\bibitem{behdad2022power}
Z.~Behdad \emph{et~al.}, ``Power allocation for joint communication and sensing in cell-free massive {MIMO},'' in \emph{2022 IEEE Global Communications Conference (GLOBECOM)}, 2022, pp. 4081--4086.

\bibitem{chu2023integrated}
Y.~Y. Chu \emph{et~al.}, ``Integrated sensing and communication in user-centric cell-free massive {MIMO} systems with {OFDM} modulation,'' in \emph{2023 IEEE 34th Annual International Symposium on Personal, Indoor and Mobile Radio Communications (PIMRC)}, 2023, pp. 1--7.

\bibitem{babu2024precoding}
N.~Babu \emph{et~al.}, ``Precoding for multi-cell {ISAC}: from coordinated beamforming to coordinated multipoint and bi-static sensing,'' \emph{IEEE Trans. Wireless Commun.}, pp. 1--1, 2024.

\bibitem{demirhan2024cellfree}
U.~Demirhan and A.~Alkhateeb, ``Cell-free {ISAC MIMO} systems: {Joint} sensing and communication beamforming,'' \emph{arXiv preprint arXiv:2301.11328}, 2024.

\bibitem{mao2023beamforming}
W.~Mao \emph{et~al.}, ``Beamforming design in cell-free massive {MIMO} integrated sensing and communication systems,'' in \emph{2023 IEEE Global Communications Conference (GLOBECOM)}, 2023, pp. 546--551.

\bibitem{behdad2024isac}
Z.~Behdad \emph{et~al.}, ``Multi-static target detection and power allocation for integrated sensing and communication in cell-free massive {MIMO},'' \emph{IEEE Trans. Wireless Commun.}, pp. 1--1, 2024.

\bibitem{elfiatoure2023cell}
M.~Elfiatoure \emph{et~al.}, ``Cell-free massive {MIMO for ISAC}: Access point operation mode selection and power control,'' \emph{arXiv preprint arXiv:2310.09032}, 2023.

\bibitem{elfiatoure2024multipletarget}
------, ``Multiple-target detection in cell-free massive {MIMO}-assisted {ISAC},'' \emph{arXiv preprint arXiv:2404.17263}, 2024.

\bibitem{mao2024csregion}
W.~Mao \emph{et~al.}, ``Communication-sensing region for cell-free massive {MIMO ISAC} systems,'' \emph{IEEE Trans. Wireless Commun.}, vol.~23, no.~9, pp. 12\,396--12\,411, 2024.

\bibitem{interdonato2019scalability}
G.~Interdonato, P.~Frenger, and E.~G. Larsson, ``Scalability aspects of cell-free massive {MIMO},'' in \emph{2019 IEEE International Conference on Communications (ICC)}, 2019, pp. 1--6.

\bibitem{bjornson2020scalable}
E.~Bj{\"o}rnson and L.~Sanguinetti, ``Scalable cell-free massive {MIMO} systems,'' \emph{IEEE Trans. Commun.}, vol.~68, no.~7, pp. 4247--4261, 2020.

\bibitem{buzzi2019user}
S.~Buzzi \emph{et~al.}, ``User-centric {5G} cellular networks: {Resource allocation and comparison with the cell-free massive MIMO approach},'' \emph{IEEE Trans. Wireless Commun.}, vol.~19, no.~2, pp. 1250--1264, 2019.

\bibitem{interdonato2023coexistence}
G.~Interdonato \emph{et~al.}, ``On the coexistence of {eMBB} and {URLLC} in multi-cell massive {MIMO},'' \emph{IEEE O. J. Commun. Soc.}, vol.~4, pp. 1040--1059, 2023.

\bibitem{dandrea2020uav}
C.~D’Andrea \emph{et~al.}, ``Analysis of {UAV} communications in cell-free massive {MIMO} systems,'' \emph{IEEE O. J. Commun. Soc.}, vol.~1, pp. 133--147, 2020.

\bibitem{3GPP36814}
\emph{Further advancements for E-UTRA physical layer aspects}, 3GPP, Technical Report 36.814 v9.2.0 2017.

\bibitem{ngo2017cell}
H.~Q. Ngo \emph{et~al.}, ``Cell-free massive {MIMO} versus small cells,'' \emph{IEEE Trans. Wireless Commun.}, vol.~16, no.~3, pp. 1834--1850, 2017.

\bibitem{liesegang2024emf}
S.~Liesegang and S.~Buzzi, ``{EMF}-aware power control for massive {MIMO}: Cell-free versus cellular networks,'' in \emph{2024 IEEE Wireless Communications and Networking Conference (WCNC)}, 2024, pp. 1--6.

\bibitem{richards2023principles}
M.~Richards and W.~Melvin, \emph{Principles of Modern Radar: Basic Principles}, ser. Principles of Modern Radar.\hskip 1em plus 0.5em minus 0.4em\relax Institution of Engineering and Technology, 2023, no. v. 1.

\bibitem{petersen2012matrix}
K.~B. Petersen and M.~S. Pedersen, \emph{The Matrix Cookbook}.\hskip 1em plus 0.5em minus 0.4em\relax Technical University of Denmark, 2012.

\bibitem{li2008mimo}
J.~Li and P.~Stoica, \emph{MIMO Radar Signal Processing}, ser. IEEE Press.\hskip 1em plus 0.5em minus 0.4em\relax Wiley, 2008.

\bibitem{sun2017mm}
Y.~Sun, P.~Babu, and D.~P. Palomar, ``Majorization-minimization algorithms in signal processing, communications, and machine learning,'' \emph{IEEE Trans. Signal Process.}, vol.~65, no.~3, pp. 794--816, 2017.

\bibitem{cvx2020}
M.~Grant and S.~Boyd, ``{CVX}: {MATLAB} software for disciplined convex programming, version 2.2,'' \url{http://cvxr.com/cvx}, 2020.

\end{thebibliography}

\end{document}